\documentclass[8pt,journal]{IEEEtran}
\usepackage{cite}
\ifCLASSINFOpdf
   \usepackage[pdftex]{graphicx}
   \DeclareGraphicsExtensions{.pdf,.jpeg,.png}
\else
   \DeclareGraphicsExtensions{.fig}
\fi

\usepackage{amsmath}
\usepackage{epstopdf}
\usepackage{subfigure}
\usepackage[font=footnotesize, format=plain, up]{caption}
\usepackage{caption}
\usepackage{rotating}
\usepackage{array, multirow}
\usepackage[caption=false,font=scriptsize,labelfont=sf,textfont=sf]{subfig}
\usepackage{fixltx2e}
\usepackage{stfloats}
\usepackage{rotating}
\usepackage{url}
\begin{document}
\pagenumbering{gobble}
\title{An Educational Kit Based on a Modular Silicon Photomultiplier System}

\author{
    \IEEEauthorblockN{Valentina~Arosio, Massimo~Caccia\thanks{M. Caccia, V. Chmill, A. Ebolese, A. Martemiyanov, F. Risigo and R. Santoro are with the Dipartimento di Scienza e Alta Tecnologia, Universita' degli Studi dell'Insubria, 22100, Como, Italy (e-mail: massimo.caccia@uninsubria.it).}\thanks{M. Locatelli, M. Pieracci and C. Tintori are with the CAEN S.p.A., 55049, Viareggio, Italy (e-mail: m.locatelli@caen.it).}, Valery~Chmill, Amedeo~Ebolese, Marco~Locatelli, Alexander~Martemiyanov, Maura~Pieracci, Fabio~Risigo, Romualdo~Santoro and Carlo~Tintori}\\
%    \IEEEauthorblockN{Massimo~Caccia, Valery~Chmill, Amedeo~Ebolese, Alexander~Martemiyanov,}\\
 %   \IEEEauthorblockN{Fabio~Risigo, Romualdo~Santoro}\\
    
   % \IEEEauthorblockA{ {\it{Dipartimento di Scienza e Alta Tecnologia,}}\\ {\it{Universita' degli Studi dell'Insubria, 22100, Como, Italy}}\\}
    
%\thanks{M. Caccia, V. Chmill, A. Ebolese, A. Martemiyanov, F. Risigo and R. Santoro are with the Dipartimento di Scienza e Alta Tecnologia, Universita' degli Studi dell'Insubria, 22100, Como, Italy (e-mail: massimo.caccia@uninsubria.it).}\thanks{M. Locatelli, M. Pieracci and C. Tintori are with the CAEN S.p.A., 55049, Viareggio, Italy (e-mail: m.locatelli@caen.it).} \\

%    \IEEEauthorblockA{{\it{CAEN S.p.A., 55049, Viareggio, Italy}}}
    
}

%\markboth{\# $N^o$ 1199 {\it{An Educational Kit Based On a Modular Silicon Photomultiplier System}}}%
%{Shell \MakeLowercase{\textit{et al.}}: Bare Demo of IEEEtran.cls for Journals}

\maketitle

\begin{abstract}
Silicon Photo-Multipliers (SiPM) are state of the art light detectors with unprecedented single photon sensitivity and photon number resolving capability, representing a breakthrough in several fundamental and applied Science domains. An educational experiment based on a SiPM set-up is proposed in this article, guiding the student towards a comprehensive knowledge of this sensor technology while experiencing the quantum nature of light and exploring the statistical properties of the light pulses emitted by a LED.
\end{abstract}

\IEEEpeerreviewmaketitle

\section{Introduction}

\IEEEPARstart{E}{xploring} the quantum nature of phenomena is one of the most exciting experiences a physics student can live. What is being proposed here has to do with light bullets, bunches of photons emitted in a few nanoseconds by an ultra-fast LED and sensed by a state-of-the-art detector, a Silicon Photo-Multiplier (hereafter, SiPM). SiPM can count the number of impacting photons, shot by shot, opening up the possibility to apply basic skills in probability and statistics while playing with light quanta. After an introduction to the SiPM sensor technology (Section~\ref{Counting photons}), the basics of the statistical properties of the random process of light emission and the sensor related effects are introduced (Section~\ref{Photon Counting Statistics}). The experimental and data analysis techniques are described in Section~\ref{Experimental techniques.}, while results and discussions are reported in Section~\ref{RandD}.

\section{Counting photons} \label{Counting photons}

SiPMs are cutting edge light detectors essentially consisting of a matrix of photodiodes with a common output and densities up to $10^4/mm^2$. Each diode is operated in a limited Geiger-Muller regime in order to achieve gains at the level of $\approx 10^6$ and guarantee an extremely high homogeneity in the cell-to-cell response. Subject to the high electric field in the depletion zone,  initial charge carriers generated by an absorbed photon or by  thermal effects trigger an exponential charge multiplication  by impact ionization, till when the current spike across the quenching resistance induces a drop in the operating voltage, stopping the process \cite{Dolgoshein, Renker:2006,Piemonte(2006)}. SiPM can be seen as a collection of  binary cells, providing altogether an information about the intensity of the incoming light by counting the number of fired cells. \\

\begin{figure}[hf]
\centering
 \includegraphics[width=0.4\textwidth]{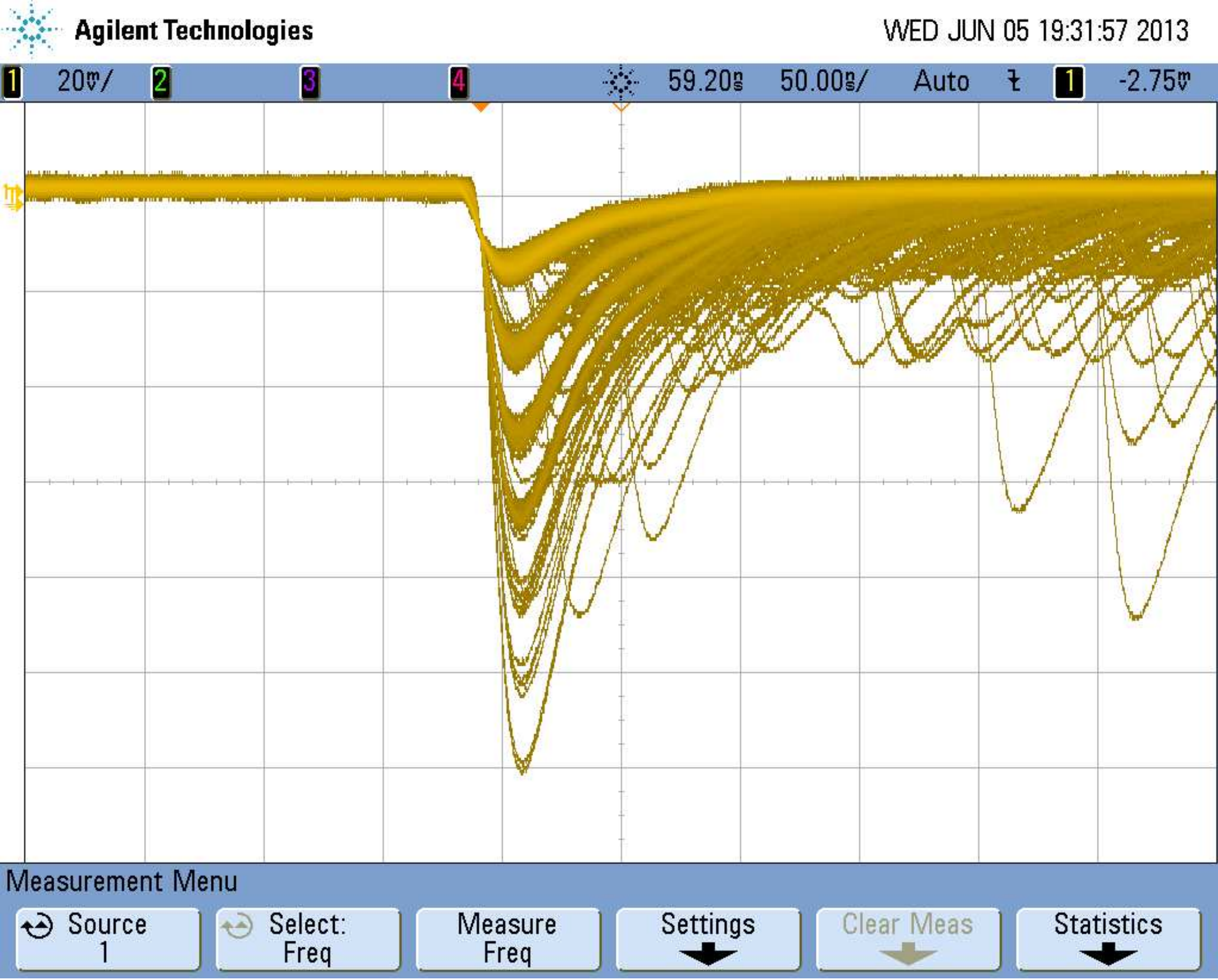}
 \caption{Response of a SiPM Hamamatsu MPPC S10362-11-100C illuminated by a light pulse.}
  \label{fig:oscilloscope}
\end{figure}

Fig.~\ref{fig:oscilloscope} shows the typical response by a SiPM to a light pulse: traces correspond to different numbers of fired cells, proportional to the number of impinging photons. Because of the high gain compared to the noise level, traces are well separated, providing a photon number resolved detection of the light field.

\vspace{-5pt}

\begin{figure}
\includegraphics[width=0.49\textwidth]{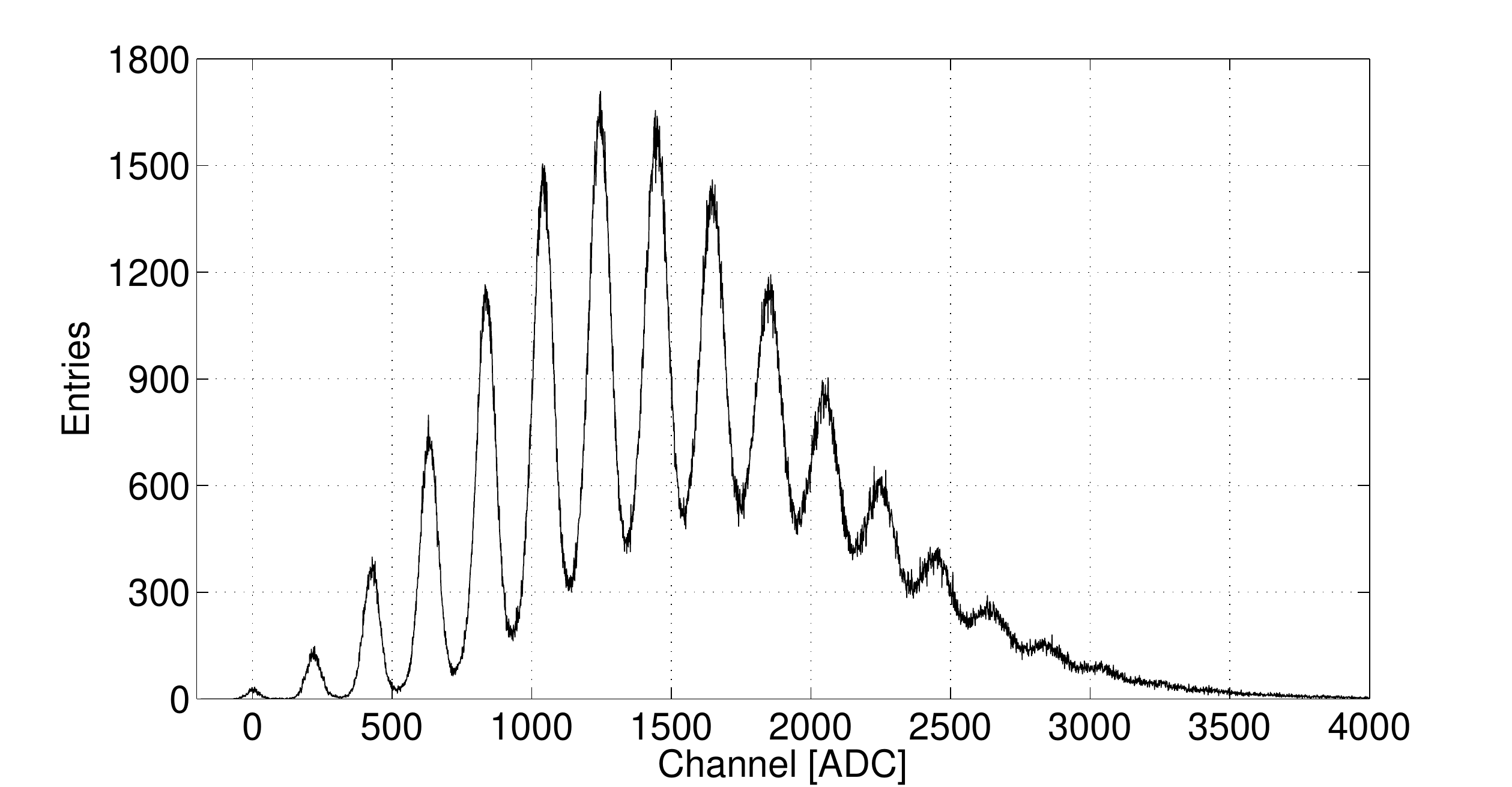}
\caption{Photoelectron spectrum probing a LED source measured with a Hamamatsu MPPC S10362-11-100C at a bias voltage of $70.3V$ and temperature of $25^o C$.}
  \label{fig:ExemplaryHistogram}
  \vspace{-15pt}
\end{figure}

This is also shown in Fig.~\ref{fig:ExemplaryHistogram}, displaying the spectrum of the SiPM response to a high statistics of pulses: every entry corresponds to the digitized released charge, measured integrating the electrical current spike during a pre-defined time interval. The peaks correspond to different number of cells fired at the same time. Each peak is well separated and occurs with a probability linked at first order to the light intensity fluctuations. An analysis of the histogram is revealing other significant characteristics:\\

\begin{itemize}
	\item The peak at $0$ corresponds to no detected photons and its width measures the noise of the system, i.e. the stochastic fluctuations in the output signal in absence of any stimulus. In the displayed histogram, $\sigma_0$~=~29$\pm1$ ADC channels.\\
	\item The peak at 1 detected photon has a width $\sigma_1$~=~38.1$\pm$0.4 ADC channels, by far exceeding $\sigma_0$. The extra contribution may be related to the fact that not all of the cells were born equal. In SiPM the homogeneity of the response is quite high \cite{2010NIMPA.620..217E, 2006NIMPA.567...57M}, however, since fired cells are randomly distributed in the detector sensitive area residual differences in the gain become evident broadening the peak.\\
	\item As a  consequence the peak width is increasing with the number N of fired cells with a growth expected to follow a $\sqrt N$ law, eventually limiting the maximum number {\it{M}} of resolved peaks.\\
\end{itemize}

\vspace{-5pt}
\begin{figure}[hf]
\centering
 \includegraphics[width=0.46\textwidth]{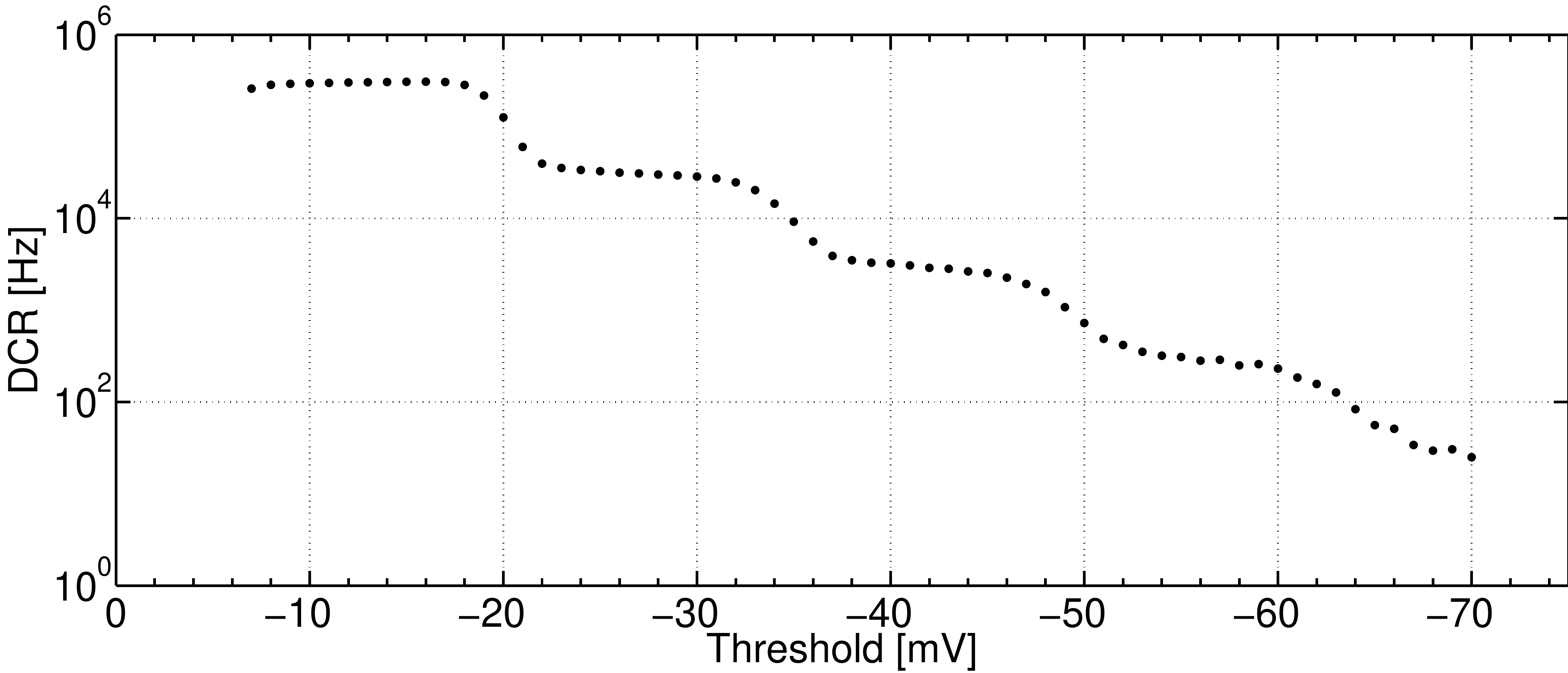}
  \caption{Measurement of the DCR of the SiPM performed at $25^o C$.}
  \label{fig:staircase}
  \vspace{-5pt}
\end{figure}
 
The detector working conditions can be optimized to maximize {\it{M}}, properly tuning the bias voltage $V_{bias}$ and balancing competing effects. On one hand, the peak-to-peak distance is linked to the single cell gain and it is expected to grow linearly with the over-voltage as:

\begin{equation}
Gain = \frac{C~\Delta V}{q_e},
\end{equation}
\\
where $\Delta V = V_{bias} - V_{Breakdown}$, C is the diode capacitance of the single cell and $q_e$ the electron charge \cite{Renker:2006}. Effects broadening the peaks may grow faster dumping the expected resolution. Among these effects it is worth mentioning Dark Counts, Optical Cross-Talk and After-Pulsing:\\

\begin{itemize}
	\item Free charge carriers may also be thermally generated. Results are spurious avalanches (Dark Counts) occurring randomly and independently from the illumination field. The Dark Count Rate (DCR) does depend from several factors: substrate, processing technology, sensor design and operating temperature \cite{Piemonte(2006)}. The over-voltage has an impact since the junction thickness volume grows with it  together with  the triggering probability, namely the probability that a charge carrier develops an avalanche \cite{Piemonte(2006)}, \cite{2010NIMPA.620..217E}. The DCR can be measured in different ways. A {\it{Stair Case Plot}} is presented in Fig.~\ref{fig:staircase} where the output from a sensor is compared to the threshold of a discriminator and the rate with which the threshold is exceeded is counted. A typical DCR is about $0.5~MHz/mm^2$.\\
	\item Dark Counts may be considered as statistically independent. However, optical photons developed during an avalanche have been shown to trigger secondary avalanches \cite{Piemonte(2006)} involving more than one cell into spurious pulses. This phenomenon is named Optical Cross-Talk (OCT). The OCT is affected by the sensor technology \cite{Piemonte(2006)},\cite{2008NIMPA.596..396D},\cite{Buzhan:2001xq},\cite{Ramillietal(2010)} and strongly depends on the bias voltage increasing the triggering probability and the gain forming the optical photon burst. The OCT can be measured by the ratio of the Dark Counts frequencies for pulses exceeding the 0.5 and 1.5 levels of the single cell amplitude, namely:

	\begin{equation}
	OCT = \frac{\nu_{1.5pe}}{\nu_{0.5pe}}.  
	\end{equation}
\\
The OCT typically ranges between 10\% and 20\% \cite{2010NIMPA.620..217E}, \cite{Ramillietal(2010)}, \cite{2009PhRvA..79d3830A}.\\
	\item Charge carriers from an initial avalanche may be trapped by impurities and released at later stage resulting in delayed avalanches named After-Pulses. For the detectors in use here, an After-Pulse rate at about the 25\% level has been reported for an overvoltage $\Delta V~=~1~V$, with a linear dependence on $V_{bias}$ and a two-component exponential decay time of 15 $ns$ and 80 $ns$ \cite{2010NIMPA.620..217E}.\\
\end{itemize}

Dark Counts, Optical Cross-Talk and After-Pulses occur stochastically and introduce fluctuations in the multiplication process that contribute to broaden the peaks in the spectrum. An exhaustive study of this effect, also known as Excess Noise Factor (ENF), exceeds the goals of this work and will not be addressed here (see for example \cite{Renker:2006}, \cite{2006NIMPA.567...57M}, \cite{Buzhan:2001xq} and \cite{2012NIMPA.695..247V}). However, the resolving power that will be introduced in the following may be considered a figure of merit accounting for the ENF and measuring the ability to resolve the number of detected photons. 

\vspace{4pt}
\section{Photon Counting Statistics} \label{Photon Counting Statistics}

Spontaneous emission of light results from random decays of excited atoms. Occurrences may be considered statistically independent, with a decay probability within a time interval $\Delta t$ proportional to $\Delta t$ itself. Being so, the statistics of the number of photons emitted within a finite time interval T is expected to be Poissonian, namely:

\begin{equation}
	P_{n,\mathrm{ph}} = \frac{\lambda^n e^{-\lambda}}{n!},
	\label{eq:PoissonPDF}
\end{equation}
\\
where $\lambda$ is the mean number of emitted photons.

The detection of the incoming photons has a stochastic nature as well, at the simplest possible order governed by the Photon Detection Probability (PDE) $\eta$, resulting in a Binomial probability to detect $d$ photons out of $n$: 

\begin{equation}
 B_{d,n}(\eta)=\left(
 \begin{array}{c}n\\d\end{array}\right)
 \eta^d (1-\eta)^{n-d}\ .\label{eq:binomial}
\end{equation}

As a consequence, the distribution $P_{d,\mathrm{el}}$ of the number of detected photons is linked to the distribution $ P_{n,\mathrm{ph}}$ of the number of generated photons by

\begin{equation}
 P_{d,\mathrm{el}} = \sum_{n=d}^{\infty}B_{d,n}(\eta) P_{n,\mathrm{ph}}\ .\label{eq:phel}
\end{equation}

However, the photon statistics is preserved and $P_{d,\mathrm{el}}$ is actually a Poissonian distribution of mean value $\mu=\lambda \eta$ \cite{Ramillietal(2010)}, \cite{2009PhRvA..79d3830A}. For the sake of completeness, the demonstration is reported in Appendix A.

Detector effects (especially OCT and After-Pulses) can actually modify the original photo-electron probability density function, leading to significant deviations from a pure Poisson distribution. Following \cite{Ramillietal(2010)} and \cite{2009PhRvA..79d3830A}, OCT can be accounted for by a parameter $\epsilon_{XT}$, corresponding to the probability of an avalanche to trigger a secondary cell. The probability density function of the number of fired cells, the random discrete variable $m$, can be written at first order as: 

\begin{equation}
P\otimes B = \sum^{floor(m/2)}_{k=0} B_{k,m-k}(\epsilon_{XT})P_{m-k}(\mu),
  \label{eq:conv}
\end{equation}
\\
where {\it {floor}} rounds $m$/2 to the nearest lower integer and $B_{k,m-k}(\epsilon_{XT})$ is the binomial probability for $m-k$ cells fired by a photon to generate $k$ extra hit by OCT. $P\otimes B$ is characterized by a mean value and variance expressed as:

\begin{equation}
\bar{m}_{P\otimes B}= \mu (1+\epsilon_{XT}) \quad \quad \sigma^2_{P\otimes B} = \mu (1+\epsilon_{XT}) . 
\end{equation}

In order to perform a more refined analysis, the probability density function of the total number of detected pulses can be calculated taking into account higher order effects \cite{Vinogradov}. The result is achieved by assuming that every primary event may produce a single infinite chain of secondary pulses with the same probability $\epsilon_{XT}$. Neglecting the probability for an event to trigger more than one cell, the number of secondary hits, described by the random discrete variable $k$, follows a geometric distribution with parameter $\epsilon_{XT}$:

\begin{equation}
  G_k(\epsilon_{XT})={\epsilon_{XT}}^k(1-\epsilon_{XT}) \qquad for \quad k=0,1,2,3... .
  \label{eq:geom}
\end{equation}

The number of primary detected pulses is denoted by the random discrete variable $\emph{d}$ and belongs to a Poisson distribution with mean value $\mu$. As a consequence, the total number of detected pulses $\emph{m}$ represents a compound Poisson process given by:

\begin{equation}
  m=\sum_{i=1}^d(1+k_i) .
\end{equation}

Then, the probability density function of $m$ is expressed as a compound Poisson distribution:  

\begin{equation}
  P\otimes G=\frac{e^{-\mu}\sum_{i=0}^m B_{i,m}\mu^i(1-\epsilon_{XT})^i {\epsilon_{XT}}^{m-i}}{m!} ,
  \label{eq:pg}
\end{equation}
\\
where
\begin{equation*}
  B_{i,m} =
\begin{cases}
1 & \text{if $i=0$ and $m=0$} \\
0 & \text{if $i=0$ and $m>0$} \\
\frac{m!(m-1)!}{i!(i-1)!(m-i)!}  & \text{otherwise}
\end{cases}
\end{equation*}
\\

The mean value and the variance of the distribution are respectively given by:

\begin{equation}
  \bar{m}_{P\otimes G}=\frac{\mu}{1-\epsilon_{XT}} \quad \sigma^2_{P\otimes G}=\frac{\mu(1+ \epsilon_{XT})}{(1-\epsilon_{XT})^2} .
  \label{eq:pgmv}
\end{equation}

These relations can be calculated referring to the definition of the probability generating function and exploiting its features \cite{Vinogradov}. The full demonstration is available in Appendix B.
\vspace{5pt}
\section{Experimental techniques} \label{Experimental techniques.}

In this section the experimental set-up and the analysis methods are presented. The optimization of the working point of a SiPM is addressed together with the recorded spectrum analysis technique.

\begin{center}
  \includegraphics[width=90mm]{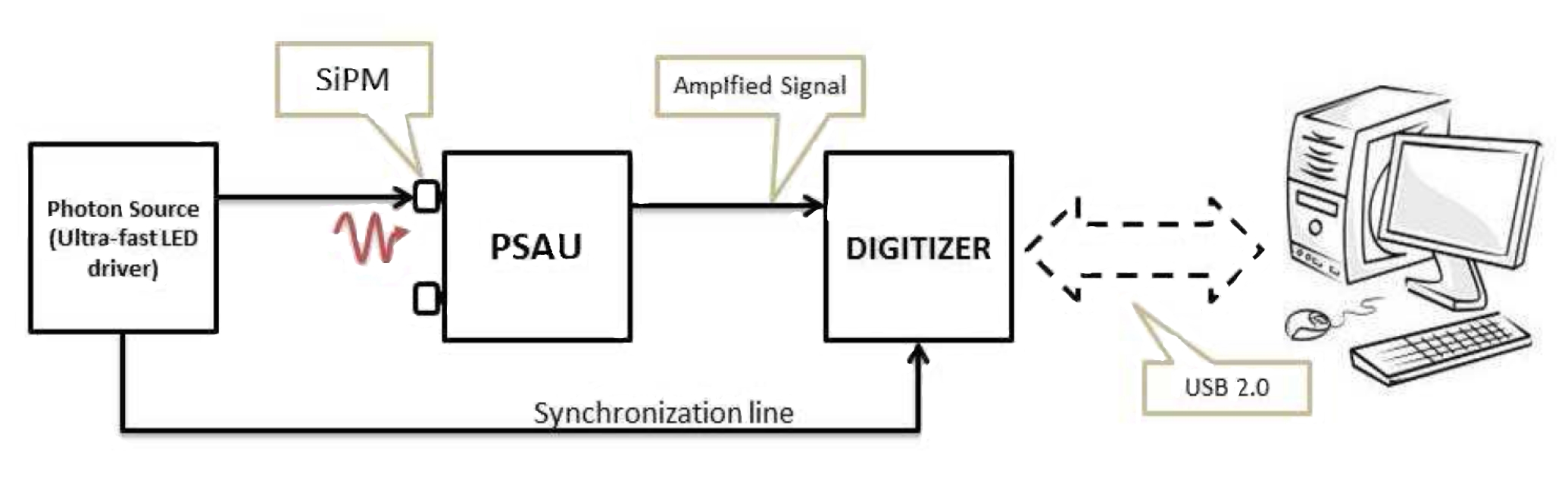}
 \captionof{figure}{Schematic layout of the experimental set-up.}
  \label{fig:BlockDiagram}
\end{center}

\subsection{Set-up and measurements}

The experimental set-up is based on the CAEN Silicon Photomultiplier Kit. The modular plug and play system contains:\\

\begin{itemize}
	\item The Two channel SP5600 CAEN Power Supply and three-stage Amplification Unit (PSAU) \cite{PSAU}, with SiPM embedding head unit. The PSAU integrates a leading edge discriminator per channel and coincidence logic.\\
	\item The two channels DT5720A CAEN Desktop Digitizer, sampling the signal at 250~MS/s over a 12 bit dynamic range. The available firmware enables the possibility to perform  charge Integration (DPP-CI), pulse shape discrimination (DPP-PSD) and advanced triggering \cite{DT5720A}.\\
	\item The ultra-fast LED (SP5601 \cite{LEDDriver}) driver emitting pulses at 400 $nm$ with FWHM of 14 $nm$. Pulses are characterized by an exponential time distribution of the emitted photons with a rising edge at sub-nanosecond level and a trailing edge with $\tau \approx$ 5 $ns$. The driver is also providing a synchronization signal in NIM standard. \\
\end{itemize}

In the current experiments the SiPM that was used is a Multi Pixel Photon Counter (MPCC) S10362-11-100C produced by HAMAMATSU Photonics\footnote{\url{http://www.hamamatsu.com/}.} (see Table~\ref{tab:hama}).

\begin{table}[h]
\renewcommand{\arraystretch}{0.95}
\caption{Main characteristics of the SiPM sensor (Hamamatsu MPPC S10362-11-100C) at a temperature of $25^o$C}
\label{tab:hama}
\centering
\begin{tabular}{lcccccl}
\hline\hline
&&&&&&\\
Number of Cells:	&$~$&~&~&~&~&	100 \\
Area:				&$~$&~&~&~&~&	$1\times1\:mm^2$ \\
Diode Dimension:	&$~$&~&~&~&~&	$100~\mu m~\times~100~\mu m$ \\
Breakdown Voltage:	&$~$&~&~&~&~&	69.2V \\
DCR:                &$~$&~&~&~&~&	600 kHz at 70.3V \\
OCT:                &$~$&~&~&~&~&	20\% at 70.3V \\
Gain:				&$~$&~&~&~&~&	$3.3\times10^6$ at 70.3V \\
PDE ($\lambda=440nm$):		&$~$&~&~&~&~&	75\% at 70.3V \\
&&&&&&\\
\hline\hline
\end{tabular}
\vspace{10pt}
\end{table}

The block diagram of the experimental set-up is presented in Fig.\ref{fig:BlockDiagram} with light pulses conveyed to the SiPM sensor by an optical fiber.

The area of the digitized signal is retained as a figure proportional to the total charge generated by the SiPM in response to the impacting photons. The integration window (or \textit{gate}) is adjusted to match the signal development and it is synchronized to the LED driver pulsing frequency.\\

\begin{figure}[h!]
\hspace{-5pt}
 \includegraphics[width= 90mm, height=35mm]{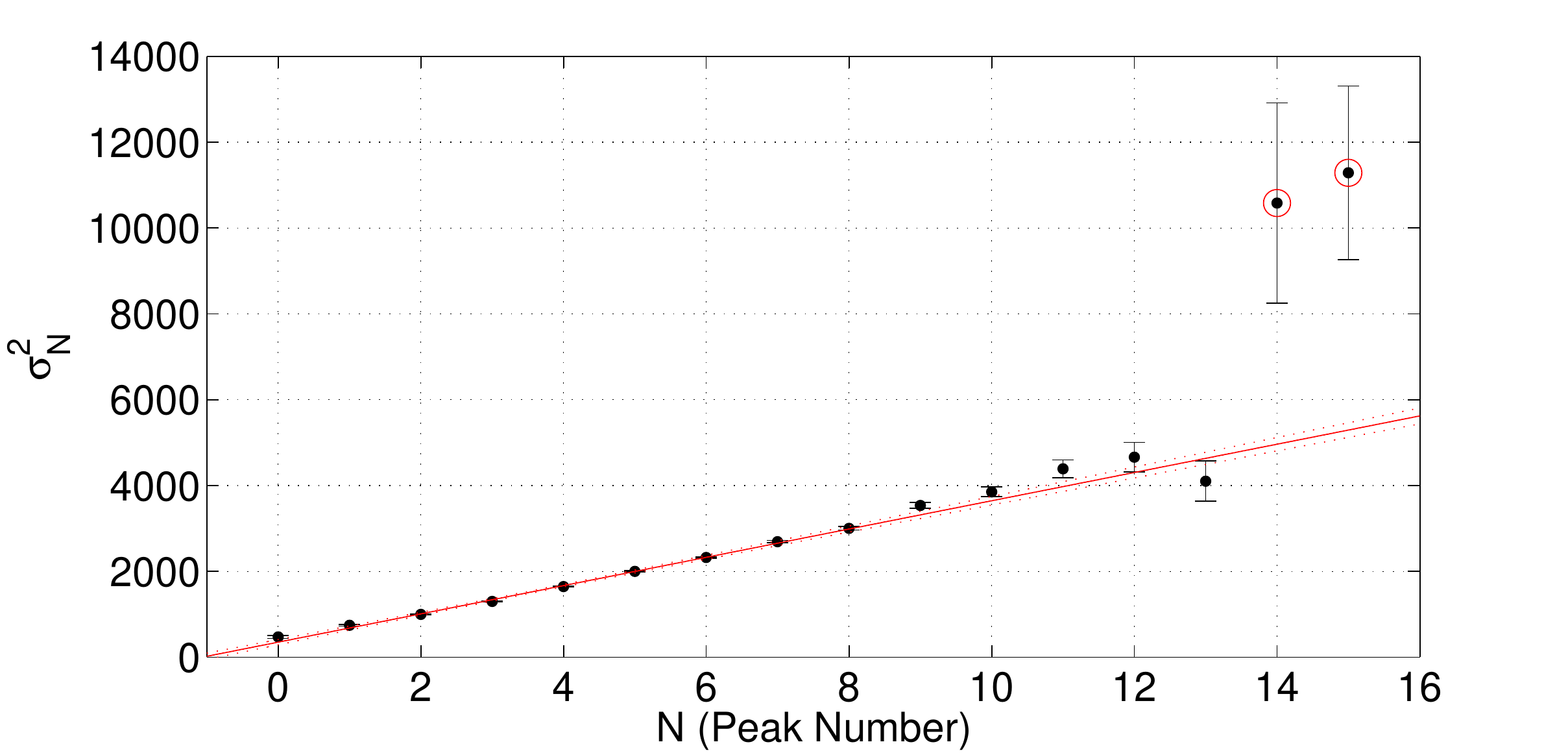}
 \caption{Peaks width for the spectrum Fig.\ref{fig:ExemplaryHistogram} by a multi-Gaussian fit. The dash lines represent the 95\% C.L. for the fit, shown with the solid line. The circles indicate the outliers.}
  \label{fig:Sigma2VsN}
\end{figure}

The proposed experimental activities start with the optimal setting of the sensor bias voltage, defined maximizing the resolving power defined as:

\begin{equation}
	R=\frac{\Delta_{pp}}{\sigma_{gain}} ,
	\label{eqn:resolutionEquation}
\end{equation}

where $\Delta_{pp}$ is the peak-to-peak distance in the spectrum and $\sigma_{gain}$ accounts for the single cell gain fluctuations:

\begin{equation}
  \sigma_{gain} = (\sigma_{1}^2-\sigma_{0}^2)^{1/2},
\end{equation}

being $\sigma_{0,1}$ the standard deviations of the 0- and 1-photoelectron peaks \cite{2011NIMPA.656...69V}. R is a figure of merit measuring the capability to resolve neighboring peaks in the spectrum. In fact, following the Sparrow criterion \cite{1961Ronchi} according to which two peaks are no longer resolved as long as the dip half way between them ceases to be visible in the superposed curves, the maximum number $N_{max}$ of  identified peaks is given by:

\begin{equation}
  N_{max} < \frac{R^2}{4},
\end{equation}

where it has been assumed the width of the peaks to grow as the squared root of the number of cells (as confirmed by the data reported in Fig.~\ref{fig:Sigma2VsN}).

The outliers, the data points that are statistically inconsistent with the rest of the data, are identified with the Thompson Tau method \cite{TTM} and discarded. 

\begin{figure}[h!]
\vspace{5pt}
\hspace{15pt}
  \includegraphics[width=75mm]{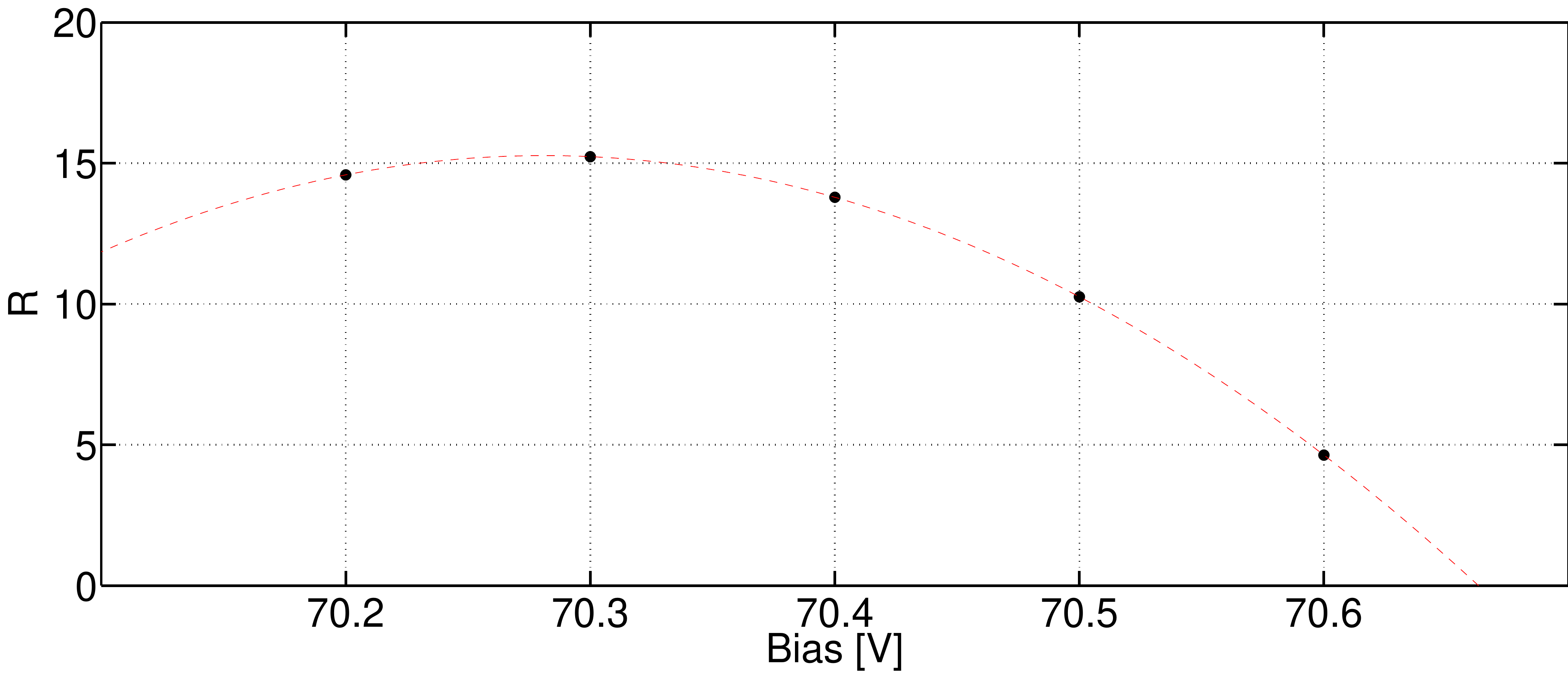}
 \caption{Scan of the resolution power R as a function of the bias voltage at fixed temperature (25$^o$C) and light intensity. The working point is given by a polynomial fit and equal to $70.28\:V$.}
  \label{fig:ResolutionPlot}
\end{figure}

A typical plot of the resolving power versus the bias voltage is presented in Fig.~\ref{fig:ResolutionPlot}. The optimal biasing value corresponds to the maximum resolution in the plot and it is used as a working point. After the sensor calibration, spectra for different light intensities are recorded and analyzed as reported below.

\begin{table}[h!]
\vspace{-15pt}
\caption{Acquisition parameters for the reference run presented in this work.}
\centering
\scalebox{0.95}{
\begin{tabular}{ccccc}
\hline\hline
					&&							&							&		\\
$V_{bias}$[V]		&&	GateWidth [ns]			& Trigger frequency [kHz]	&	Temperature [$°C$]				 \\
					&&							&							&		\\
$70.3$				&&	300						&	100  					&	25.0								\\
					&&							&							&		\\
\hline\hline
\end{tabular}
}
\end{table}

\subsection{Multi-peak spectrum analysis} \label{Data analysis techniques }

Spectra recorded  in response to  photons impacting on the sensor can be seen as a superposition of Gaussians, each corresponding to a well defined number of fired cells. The key point in the analysis technique is the estimation of the area underneath every peak, allowing the reconstruction of the probability density functions.

Initially, areas can be estimated by a {\it{Pick\&Play}} (hereafter, P\&P) procedure on the spectrum. In fact, a binned Gaussian distribution of $N_{pk}$ events may be written as:  

\begin{equation}
    y_i = y(x_i) = y_{max}e^{-\frac{(x_i-x_0)^2}{2\sigma^2}},
\end{equation}
\\
where $y(x_i)$ is the number of events in the bin centered on $x_i$ and $y_{max}$ is the peak value, measured in $x_0$. Since $y_{max} = N_{pk}/(\sigma \sqrt{2\pi})$, knowing the content of the bin centered in $x_0$ and estimating $\sigma$ leads to $N_{pk}$. The standard deviation can also be calculated in a simple way by the Full Width at Half Maximum ($FWHM$), obtained searching for the position of the bins with a content equals to $y_{max}/2$ and presuming that $FWHM = 2.355\times \sigma$. Advantages and limitations of this method are quite obvious: its applicability is straightforward and essentially requires no tool beyond a Graphical User's Interface ({\it{GUI}}) for the control of the set-up; on the other hand, it can be applied only to peaks with a limited overlap and uncertainties can only be obtained by repeating the experiment. In order to overcome these limitations, a {\it{Multi-Gaussian Fit}} (MGF) procedure was implemented in MATLAB to analyze the full spectrum, according to the following work flow:\\

\begin{itemize}
	\item {\bf {Initialization}}. Robustness and efficiency of minimization algorithms is guaranteed by having an educated guess of the parameter values and by defining boundaries in the parameter variation, a procedure increasingly important as the number of parameters grow. Initial values are provided in an iterative procedure:\\
\vspace{5pt}
	\begin{itemize}
	\item The user is required to identify by pointing \& clicking on the spectrum the  peak values and their position for 3 neighboring Gaussians, fitted to improve the estimate.\\
	\item Initial values for every Gaussian are estimated by relying on the peak-to-peak distance from the previous step, presuming the signal from the 0-cell peak to be centered in the origin of the horizontal scale and assuming the standard deviation grows as the squared root of the number of cells.\\
	\end{itemize}
	
	\item {\bf{Fit}}. Spectra are fitted to a superposition of Gaussians with a non-linear $\chi^2$ minimization algorithm presuming binomial errors in the content of every bin. The most robust convergence over a large number of tests and conditions have been empirically found bounding parameters to vary within 20\% of the initial value for the peak position, 30\% for the area and 50\% for the standard deviation.
\end{itemize}
\vspace{5pt}
\section{Results and discussions.} \label{RandD}

\begin{figure}[h!]
\centering
        \begin{subfigure}{\includegraphics[width=0.45\textwidth]{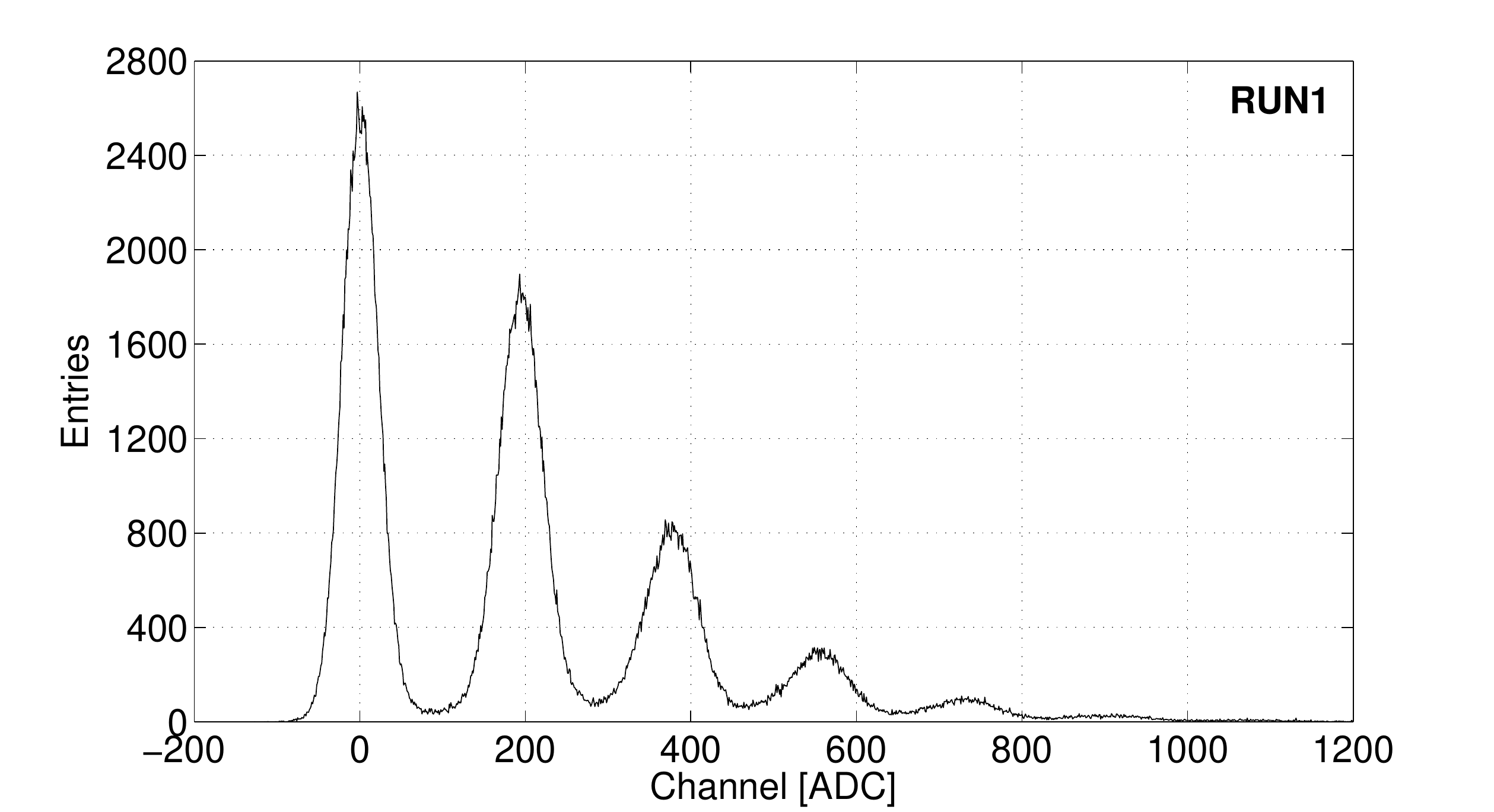}}
        \label{fig:AllRawData1}
        \vspace{-5pt}
        \end{subfigure}
                \begin{subfigure}{\includegraphics[width=0.45\textwidth]{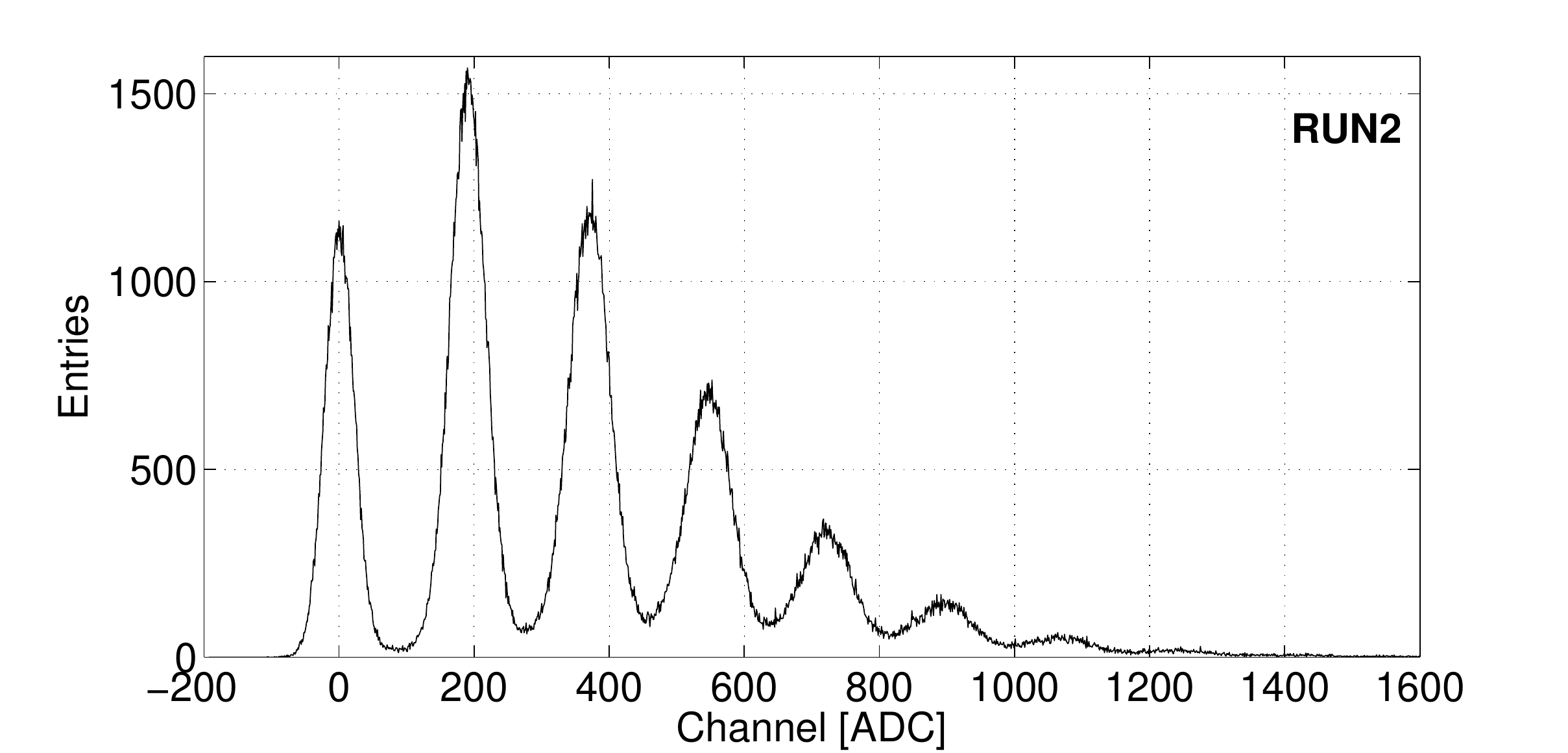}}
        \label{fig:AllRawData2}
                \vspace{-5pt}
        \end{subfigure}
       \begin{subfigure}{\includegraphics[width=0.45\textwidth]{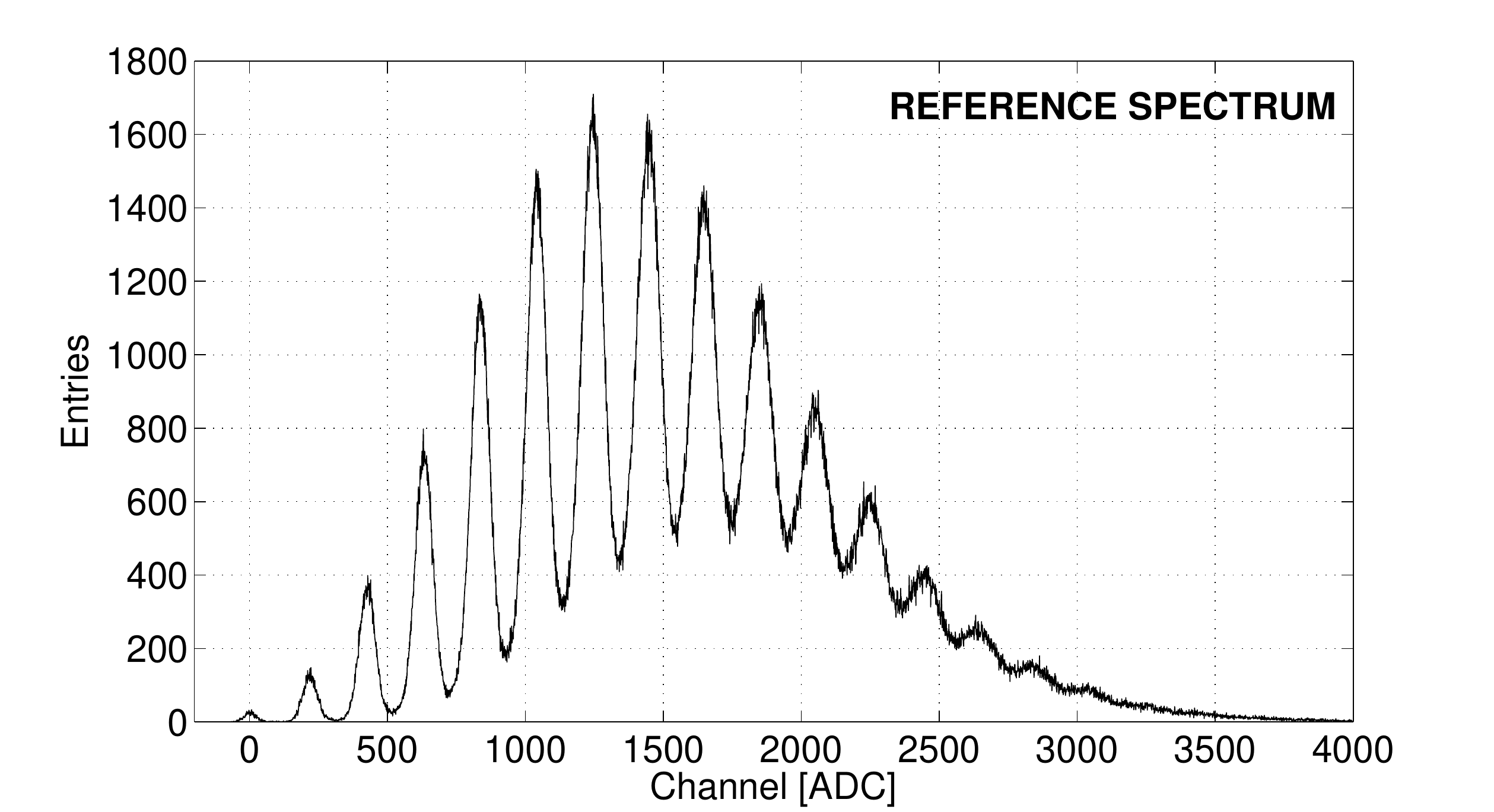}}
        \label{fig:AllRawData3}\vspace{-5pt}
        \end{subfigure}        
        \caption{Exemplary spectra. The mean number $\mu_{MI}$ of photo-electrons is measured to be $1.080\pm0.002$ (RUN1), $1.994\pm0.003$ (RUN2) and $7.81\pm0.01$ (Reference Spectrum).}
  \label{fig:ThreeInOnePlot}
 \end{figure}

Exemplary spectra for three light intensities were recorded and the raw data distributions are shown in Fig.~\ref{fig:ThreeInOnePlot}, where the horizontal scale in ADC channels measures the integrated charge in a pre-defined gate. In the following, the analysis steps are detailed for the distribution corresponding to the highest mean photon number, hereafter identified as the {\it{Reference Spectrum}}. Remaining spectra will be used to assess the robustness of the approach and the validity of the model, with the results summarized at the end of the section.

\vspace{5pt}

\begin{figure}[h!]
\vspace{-5pt}
  \includegraphics[width=90mm]{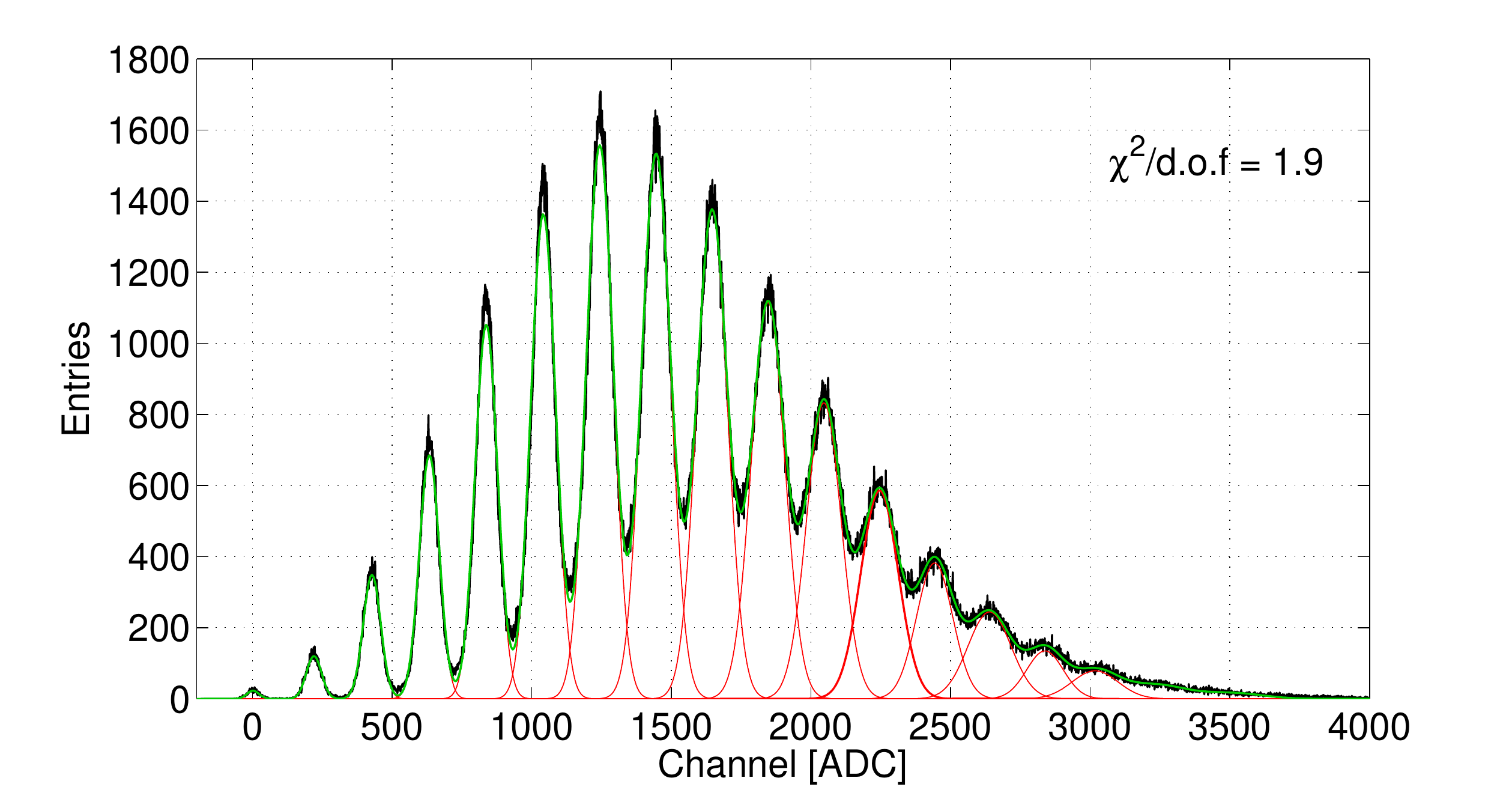}
  \caption{Outcome of the MGF procedure. Individual Gaussians are in red, while their superposition is displayed in green. The $\chi^2 / d.o.f.$ measures the fit quality.}
  \label{fig:Superposed}
\end{figure}

Spectra are seen as a superposition of Gaussians, with parameters estimated according to the methods introduced in Section~\ref{Experimental techniques.}. The outcome of the procedures for the {\it{Reference Spectrum}} is reported in Table~\ref{tab:DataElaborated} for the P\&P and the MGF procedures. For the former, uncertainties in the estimated parameters are the standard deviations from five data sets acquired in identical conditions while for the latter errors result from the fitting procedure (Fig.~\ref{fig:Superposed}).

The characteristics of the experimental distribution can initially be studied referring to the mean number of fired cells. A {\it{Model Independent}} (MI) estimate is provided by: 

\begin{equation}
	\mu_{MI} = \frac{\overline{ADC}}{\overline{\Delta_{pp}}},
		\label{eqn:meanConfront}
\end{equation}

where

\begin{equation}
\overline{ADC} =  \frac{\Sigma_i y_i ADC_i}{\Sigma_i y_i}
\label{eqn:meanConfront2}
\end{equation}
\\
is the mean value of the experimental distribution (being $y_i$ the number of events for the $i^{th}$ bin) and $\overline{\Delta_{pp}}$ is the mean peak-to-peak distance, defining the gauge to convert values in ADC channels to number of cells.

\begin{table}[h!]
\caption{Peak position, width and experimental probability of having N photo-electrons from the Pick\&Play (P\&P) procedure, compared to the results from the Multi-Gaussian Fit (MGF). The results are for the reference spectrum.}
\label{tab:DataElaborated}
\scalebox{0.77}{
\begin{tabular}{ccccccc}
\hline\hline
&&&&&& \\
  &\multicolumn{2}{c}{$PeakPosition[ADC]$}&\multicolumn{2}{c}{$PeakWidth[ADC]$}&\multicolumn{2}{c}{$Exp.~Probability$} \\
&&&&&& \\
\hline
&&&&&& \\
N &$P\&P$     &$MGF$         &     $P\&P$   &    $MGF$        & $P\&P    $   &   $MGF    $ \\
&&&&&& \\
0 & $3\pm1$   &$2.1\pm0.9$ &      $22\pm1$  &$21.7\pm0.8$     &$ 0.092\pm0.006$&      $ 0.09\pm0.01$ \\
1 & $220\pm1$ &$220.1\pm0.4$ &    $25\pm1$  &$27.3\pm0.3$     &$ 0.53\pm0.02$&       $ 0.56\pm0.01$  \\
2 & $427\pm1$ &$428.0\pm0.3$ &    $30\pm1$  &$31.5\pm0.2$     &$ 1.75\pm0.06$ &      $ 1.86\pm0.02$  \\
3 & $635\pm1$ &$633.6\pm0.2$ &    $32\pm1$  &$36.0\pm0.2$     &$ 3.8\pm0.1$  &       $ 4.17\pm0.02$   \\
4 & $838\pm2$ &$837.5\pm0.2$ &    $38\pm1$  &$40.5\pm0.2$     &$7.0\pm0.2$  &        $7.21\pm0.04$   \\
5 & $1044\pm2$&$1041.3\pm0.2$&    $41\pm1$  &$44.7\pm0.2$     &$9.9\pm0.2$  &        $10.30\pm0.04$   \\
6 & $1247\pm2$&$1243.7\pm0.2$&    $45\pm1$  &$48.2\pm0.2$     &$12.2\pm0.3$  &       $12.67\pm0.05$   \\
7 & $1449\pm3$&$1445.6\pm0.2$&    $50\pm3$  &$51.9\pm0.3$     &$13.4\pm0.8$  &       $13.43\pm0.06$   \\
8 & $1650\pm4$ &$1645.8\pm0.3$&   $57\pm2$  &$54.8\pm0.4$     &$13.3\pm0.5$      &   $12.71\pm0.07$   \\
9 & $1853\pm4$ &$1846.4\pm0.4$  & $67\pm2$  &$59.5\pm0.6$     &$12.9\pm0.4$      &   $ 11.2\pm0.1$   \\
10&     $- -$ &$2046.5\pm0.6$   & $--$      &$62.0\pm0.9$     &  $--$        &       $ 8.7\pm0.1$   \\
11&     $- -$ &$2245\pm1$    &    $--$      &$66\pm2$         &  $--$        &       $ 6.6\pm0.2$   \\
12&     $- -$ &$2445\pm1$    &    $--$      &$68\pm2$         &$--$          &       $ 4.4\pm0.2$   \\
13&     $- -$ &$2632\pm2$   &     $--$      &$65\pm3$         &  $--$        &       $ 2.4\pm0.1$   \\
&&&&&& \\\hline\hline
\end{tabular}
}
\end{table}

The value of $\mu_{MI}$ can be compared to what is estimated presuming a pure Poissonian behaviour and referring to the probability $P(0)$ of having no fired cell when the expected average value  is  $\mu_{ZP}$, where $ZP$ stands for {\it{Zero Peak}}:
 
 \begin{equation}
 \mu_{ZP}=-ln( P(0))~=-ln\left(\frac{A_{0}}{A_{tot}}\right),
 	\label{eqn:PoissP0}
 \end{equation}

being $A_{0}$ the area underneath the first peak of the spectrum and $A_{tot}$ the total number of recorded events. Results are shown in Table~\ref{tab:Res1}.

\begin{table}[!h]
\vspace{5pt}
\caption{Estimates of the mean number of fired cells by the average value of the experimental distribution and from the probability of having none, assuming an underlying Poissonian distribution. Errors result from the uncertainties in the peak-to-peak distance and in the area of the zero-cell peak.}
\label{tab:Res1}
\centering
\begin{tabular}{ccccccc}
\hline\hline
	         &&&   $\mu_{MI}$	 &&&	$\mu_{ZP}$ \\
\\
P\&P         &&& 7.6 $\pm$ 0.3    &&&      6.99 $\pm$ 0.06\\
 \\
 $MGF$       &&& 7.81 $\pm$ 0.01  &&&  7.08 $\pm$ 0.03\\
 \\
 \hline\hline
\end{tabular}
\vspace{10pt}
\end{table}

The P\&P procedure shows a good compatibility with the hypothesis, while the MGF procedure, due to the smaller errors, presents an evident discrepancy. \\

The question can be further investigated considering the full distribution and comparing the experimental probability density function with the assumed model distribution by a $\chi^2$ test, where:

\begin{equation}
	\chi^2 = \sum_{k=0}^{Npeaks-1} w_{k} \times (A_{obs,k} - A_{model,k})^2,
	\label{eqn:RedChiSquareEq}
\end{equation}
\\
being $A_{obs,k}$  the number of events in the $k^{th}$ peak of the distribution, $A_{model,k}$ the corresponding number estimated from the reference model and $w_k$ the weights accounting for the uncertainties in the content of every bin. Presuming a Poissonian distribution with mean value $\mu_{MI}$, the returned values of the $\chi^2/d.o.f.$ are $\approx 20$ for the $P\&P$ procedure and $\approx 300$ for the $MGF$. The $\chi^2/d.o.f.$ values, even assuming $\mu$ as a free parameter, exceeds the 99\%~C.L. for both methods confirming that the experimental distribution may not be adequately described by a pure Poissonian model. \\

As a further step, the spectra were compared to the $P\otimes G$ distribution model introduced in Section~\ref{Photon Counting Statistics}, Eq.~\ref{eq:pg}, where the actual number of fired cells results from  avalanches triggered by the incoming photons and by the optical cross-talk. The optimal values of the model parameters, namely the cross-talk probability $\epsilon_{XT}$ and the mean value $\mu$ of the distribution of cells fired by photons, are determined by a grid search according to the following iterative procedure \cite{Lyons:1986em} :\\
 
 \begin{itemize}
	\item  the $\chi^2/d.o.f.$ surface, henceforth referred to as $\Sigma$, is sliced with planes orthogonal to the $\epsilon_{XT}$ dimension, at values $\tilde{\epsilon}_{XT}$ changed with constant step;\\
	\item in each slice, the  minimum of the $\Sigma(\tilde{\epsilon}_{XT},\mu))$ curve is searched and the value $\mu_{min,0}$ corresponding to the minimum is identified;\\
	\item the $\Sigma(\epsilon_{XT},\mu_{min,0})$ curve is scanned and the position $\epsilon_{XT}^*$ of the minimum is identified by a local parabolic fit, to overcome the limitations by the choice of the step in the grid;\\
	\item the procedure is repeated for $\Sigma( \epsilon_{XT}^*,\mu)$ vs $\mu$, leading to the determination of the minimum in $\mu^*$.\\
\end{itemize}

\begin{figure}[h]
\centering
        \begin{subfigure}{\includegraphics[width=9.0cm, height=5.0cm]{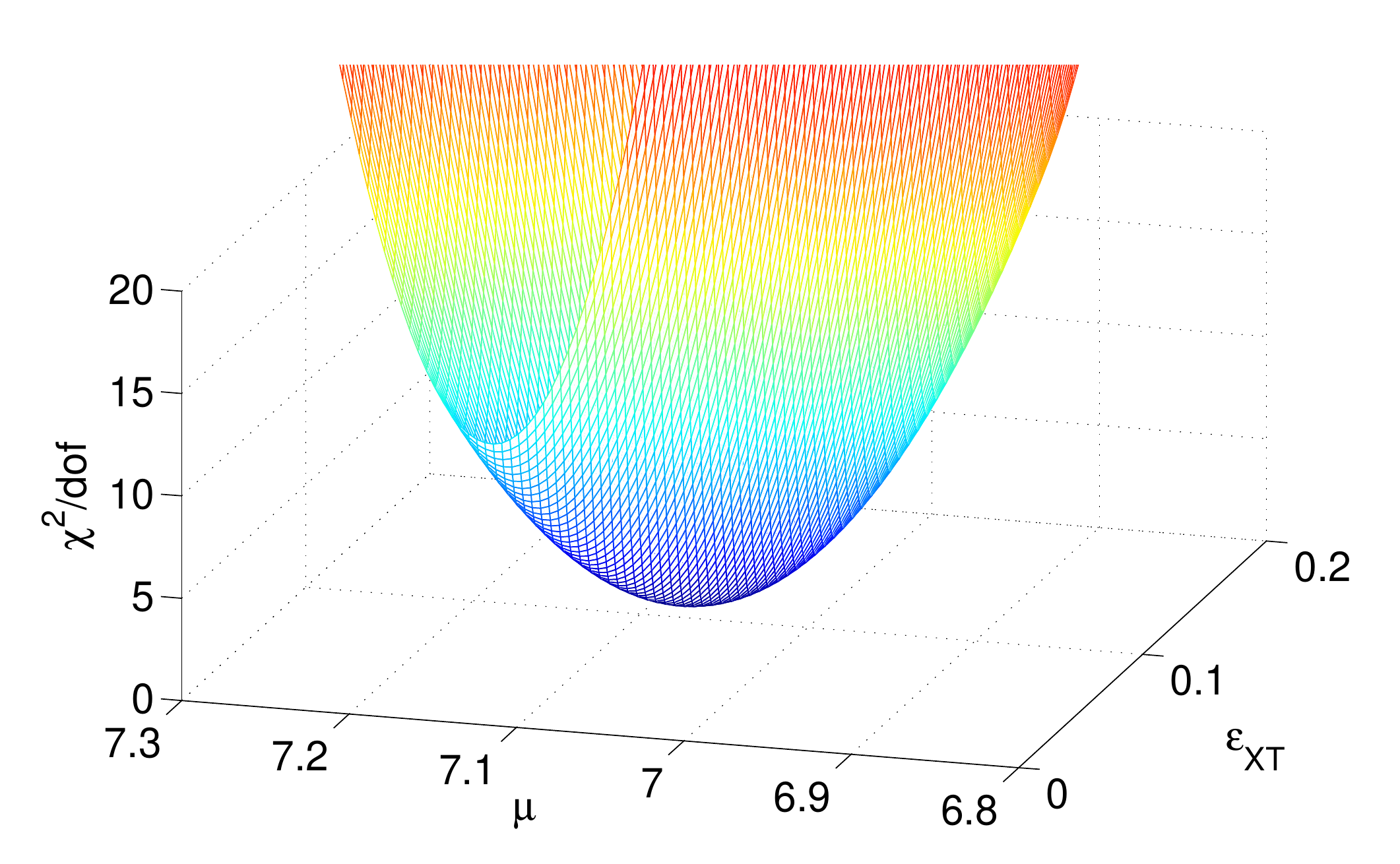}}
        \label{fig:TBD1}
        \end{subfigure}
        \vspace{-5pt}
        \begin{subfigure}{\includegraphics[width=9.0cm ,height=5.5cm]{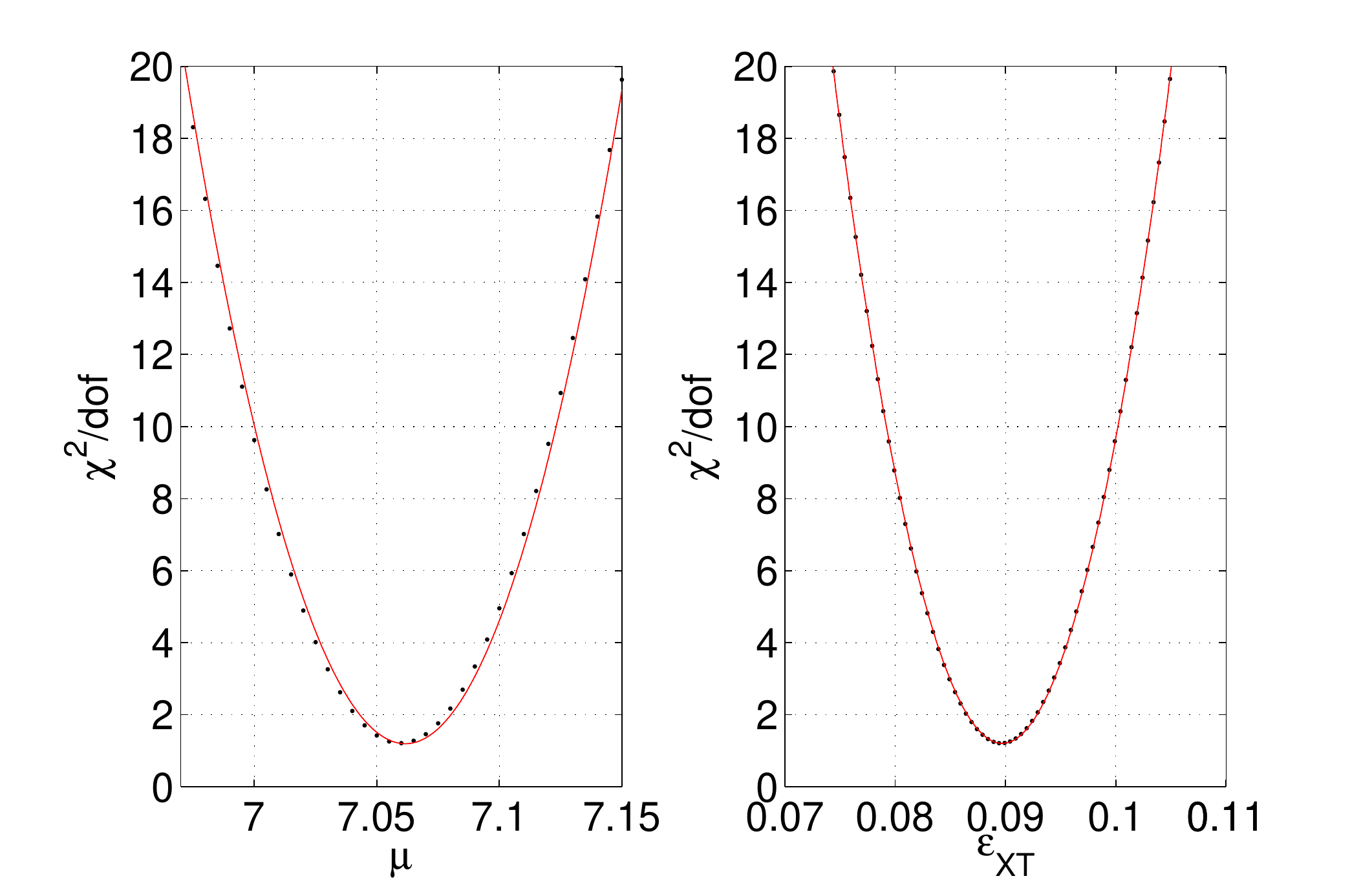}}
        \label{fig:Parabole2}
        \end{subfigure}
        \vspace{-5pt}
        \caption{$\chi^2/d.o.f.$ surface (top panel) and parabolic behavior nearby its minimum (bottom).}
  \label{fig:Parabole}
\vspace{-5pt}
\end{figure}

This method leads to estimate the optimal parameters $\mu^*$ and $\epsilon_{XT}^*$ by the minimization of the $\chi^2/d.o.f.$ surface for the two variables $\mu$ and $\epsilon_{XT}$ independently. The surface $\Sigma$ and the $\Sigma(\epsilon_{XT}^*,\mu)$ and $\Sigma(\epsilon_{XT},\mu^*)$ curves are shown in Fig.~\ref{fig:Parabole}. Uncertainties are calculated assuming a parabolic shape of the $\chi^2/d.o.f.$ curves, leading to variances estimated by the inverse of the coefficient of the quadratic term \cite{Lyons:1986em}, \cite{2003drea.book.....B}. The results for the reference spectrum are $\mu^* = 7.06\pm0.02$ and $\epsilon_{XT}^* = 0.090\pm0.004$.

In order to account for the two-parameter correlation in the calculation of the uncertainties, it is worth to referring to the \emph{confidence region} of the joint probability distribution \cite{numericalrec} \cite{glencowan}. When the parameters are estimated minimizing the $\chi^2$ distribution, confidence levels correspond to regions defined by iso-$\chi^2$ curves. For two parameters, the region assumes an elliptic shape around the $\Sigma$ minimum, $\chi^2_{\text{min}}$. The $\Sigma$ contour at the constant value of $\chi^2_{\text{min}}+1$ plays a crucial role due to its specific properties. In fact, the resulting ellipse contains $\sim$~38.5$\%$ of the joint parameter probability distribution and its projections represent the $\sim$ 68.3$\%$ of confidence interval for each parameter ($\sigma_1$ and $\sigma_2$). In addition, the correlation $\rho$ among the parameters may be written as:

\begin{equation}
\rho= \frac{\sigma_1^2-\sigma_2^2}{2\sigma_1\sigma_2}  \tan{2\theta} ,
  \label{eq:tan}
\end{equation}
\\
where $\theta$ represents the counter-clockwise rotation angle of the ellipse. The detailed demonstration is reported in Appendix C. 

\begin{figure}[h!]
\centering
 \hspace{-10pt}
  \includegraphics[width=9.0cm]{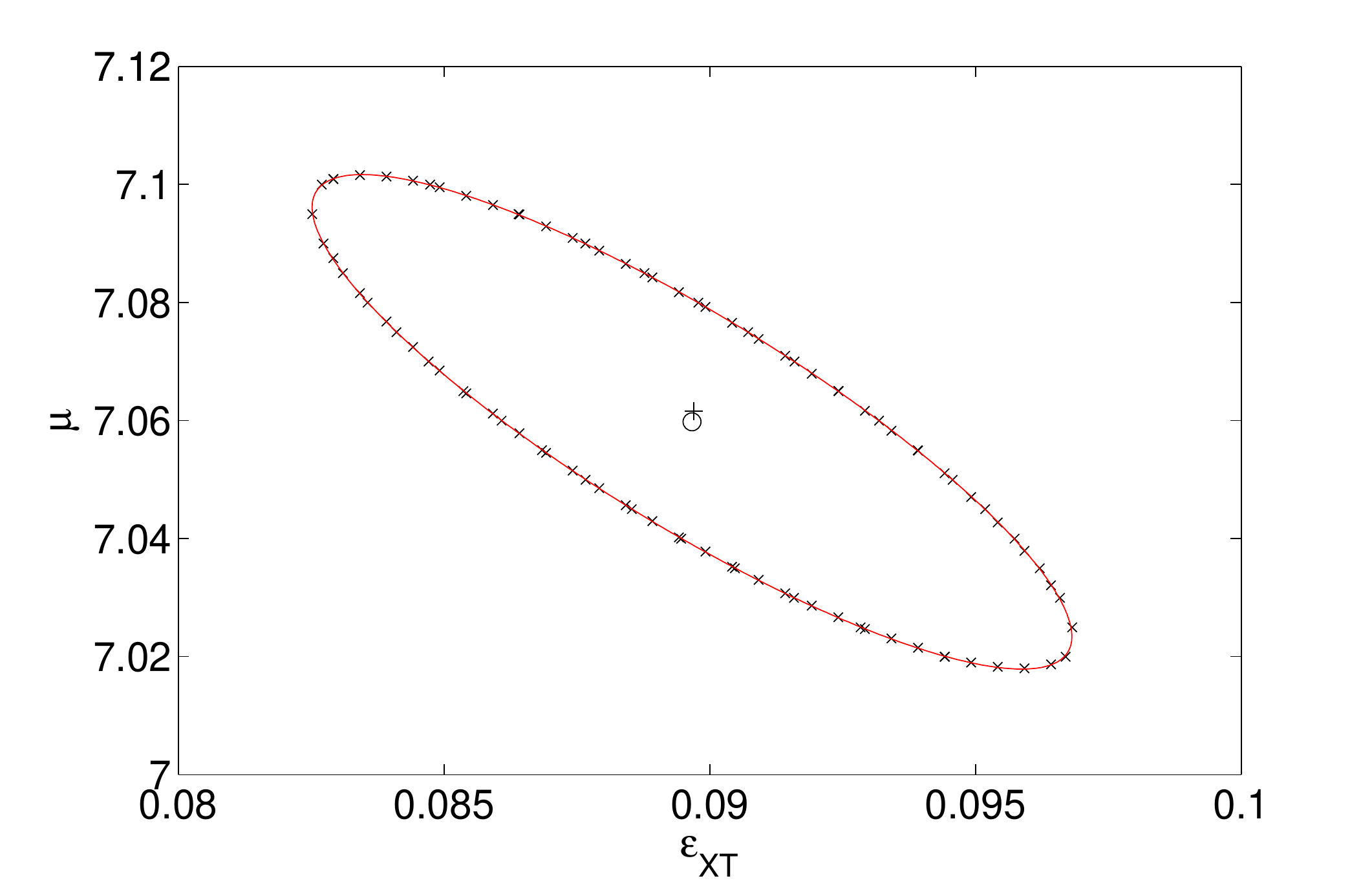}
   \caption{The point of the $\chi^2/d.o.f.$ surface at the constant value of $\chi^2_{\text{min}}+1$ are the black crosses, the fit curve is the red line, the center of the ellipse ($\mu^0$,$\epsilon_{XT}^0$) is represented with the black circle and the point ($\mu^*$,$\epsilon_{XT}^*$) is shown with the black cross.}
  \label{fig:ellipse}
  \vspace{-5pt}
\end{figure}

In this specific case, the $\chi^2_{\text{min}}$ value is determined evaluating the $\chi^2/d.o.f.$ surface at the point of coordinates ($\mu^*$, $\epsilon_{XT}^*$) while the $\Sigma$ contour at $\chi^2_{\text{min}}+1$ is shown in Fig.\ref{fig:ellipse} (black crosses). The fit curve (red line) returns the value of the ellipse center ($\mu^0$,$\epsilon_{XT}^0$) (black circle). The projections of the ellipse on the $\mu$ and $\epsilon_{XT}$ axes are the uncertainties on the two values. The results for the reference spectrum are $\mu^0$~=~7.06~$\pm$~0.05 and $\epsilon_{XT}^0 = 0.09 \pm 0.01$. Comparing these values with ($\mu^*$,$\epsilon_{XT}^*$) (black cross) it is possible to infer that the correlation does not affect the determination of the parameter central values while increases their standard deviation by a factor of about two. As a consequence, $\mu^0$ and $\epsilon_{XT}^0$ with their uncertainties are retained as the best estimate of the model parameter values.
  
The angle returned by the ellipse fit is used to calculate the correlation $\rho$ between the two parameters through the equation \eqref{eq:tan}. The result for the reference spectrum is $\rho = -0.8$. Then, applying the relation \eqref{eq:pgmv} and exploiting the full covariance matrix, the value and the uncertainty of the mean of the $P\otimes G$ model can be obtained. For the reported spectra it results to be $7.76 \pm 0.03$.

\begin{figure}[h!]
\hspace{-20pt}
  \includegraphics[width=10.0cm]{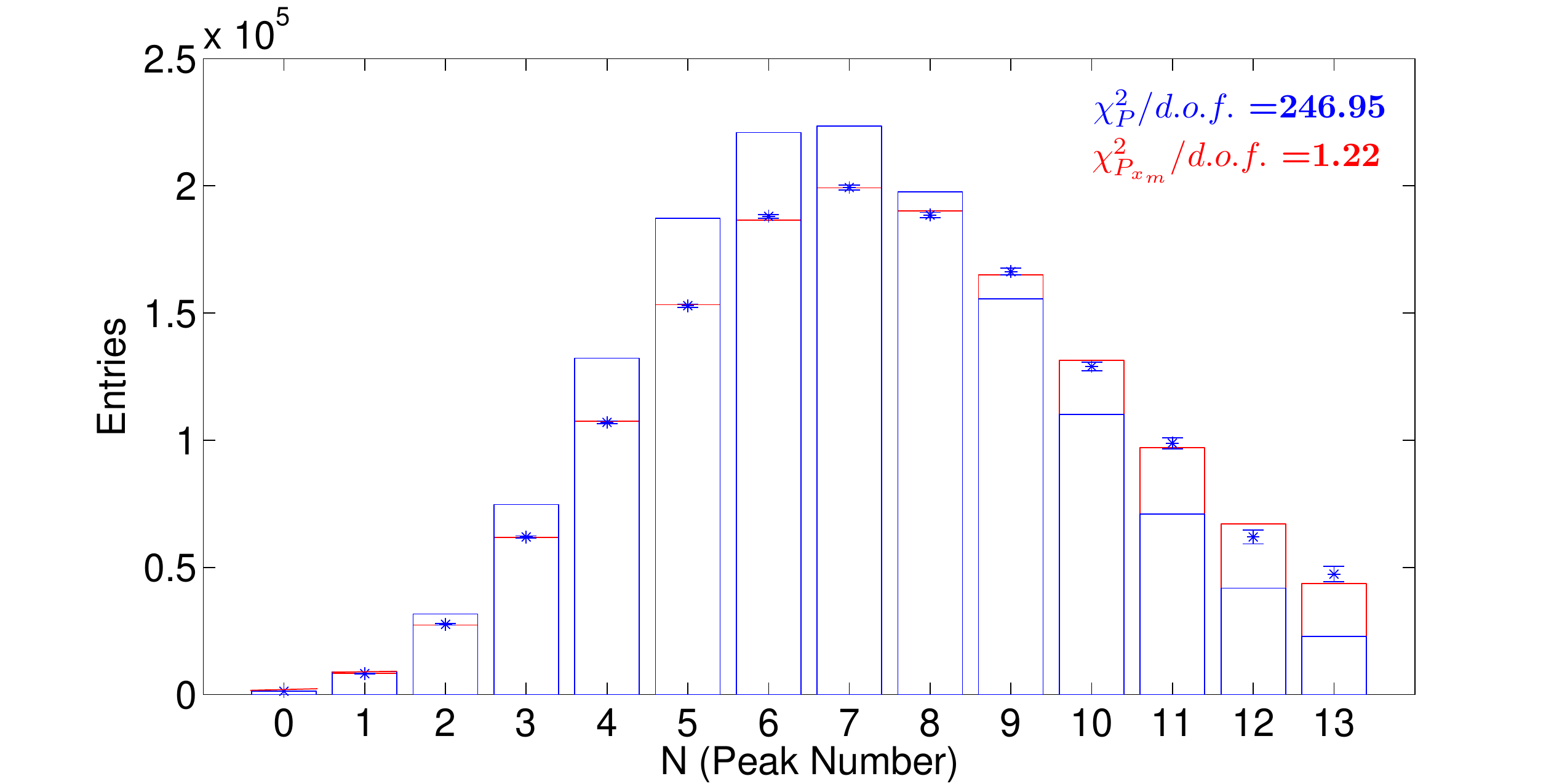}
  \caption{Data from the reference spectrum are compared to a simple Poissonian model with mean value $\mu_{ZP}$ (blue) and to the $P\otimes G$ model (red), accounting for the optical cross-talk. The $\chi^2$ value rule out the former at 99\% C.L..}
  \label{fig:FinalHisto}
\vspace{-5pt}
\end{figure}

The result of the fit to the data distribution with the $P\otimes~G$ probability function is displayed in Fig.~\ref{fig:FinalHisto}, showing an excellent agreement between data and model.

The quality of the result is confirmed by the data reported in Table~\ref{tab:Res1v1}, where the mean value of the Poissonian distribution obtained by the ellipse fit ($\mu^0$) and by the {\it Zero Peak} are compared, together with a comparison between $\mu_{MI}$ and $\mu^0/(1-\epsilon_{XT}^0)$, the mean value of the $P\otimes G$ distribution.

\newpage

\begin{table}[h!]
\vspace{-5pt}
\caption{Estimates of the mean number of fired cells for $P\otimes G$ model.}
\label{tab:Res1v1}
\centering
\scalebox{1.0}{
\begin{tabular}{p{2.6cm}ccc}
\hline\hline\\
                                          &   $\mu^0$	       &&	$\mu_{ZP}$ \\
\\
Mean Value of the Poissonian distribution & $7.06 \pm 0.05$    &&	$7.08 \pm 0.03$ \\
\\ \hline \\
                                          &   $\mu_{MI}$	   &&	$\mu^0/(1-\epsilon_{XT}^0)$ \\
\\ 
Mean Number of Fired Cells                 & $7.81 \pm 0.01$      &&  $7.76 \pm 0.03$ \\
 \\
 \hline\hline
    \end{tabular}    }
\vspace{15pt}
\end{table}

\begin{figure}[h!]
\hspace{-10pt}
        \begin{subfigure}{\includegraphics[width=10.0cm]{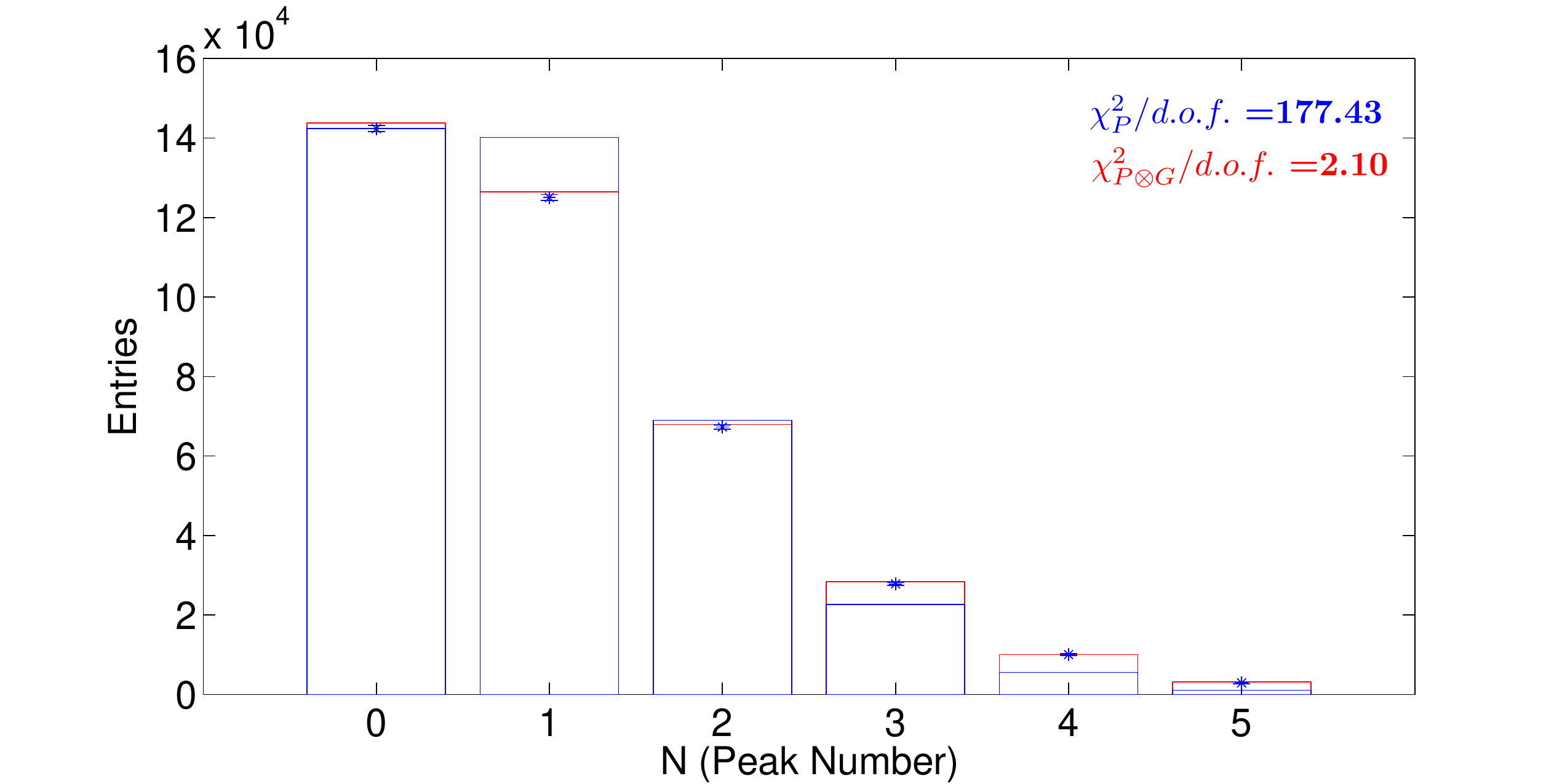}}
        \label{fig:AllRawData2}
        \end{subfigure}
\hspace{-15pt}
\vspace{5pt}        
        \begin{subfigure}{\includegraphics[width=10.0cm]{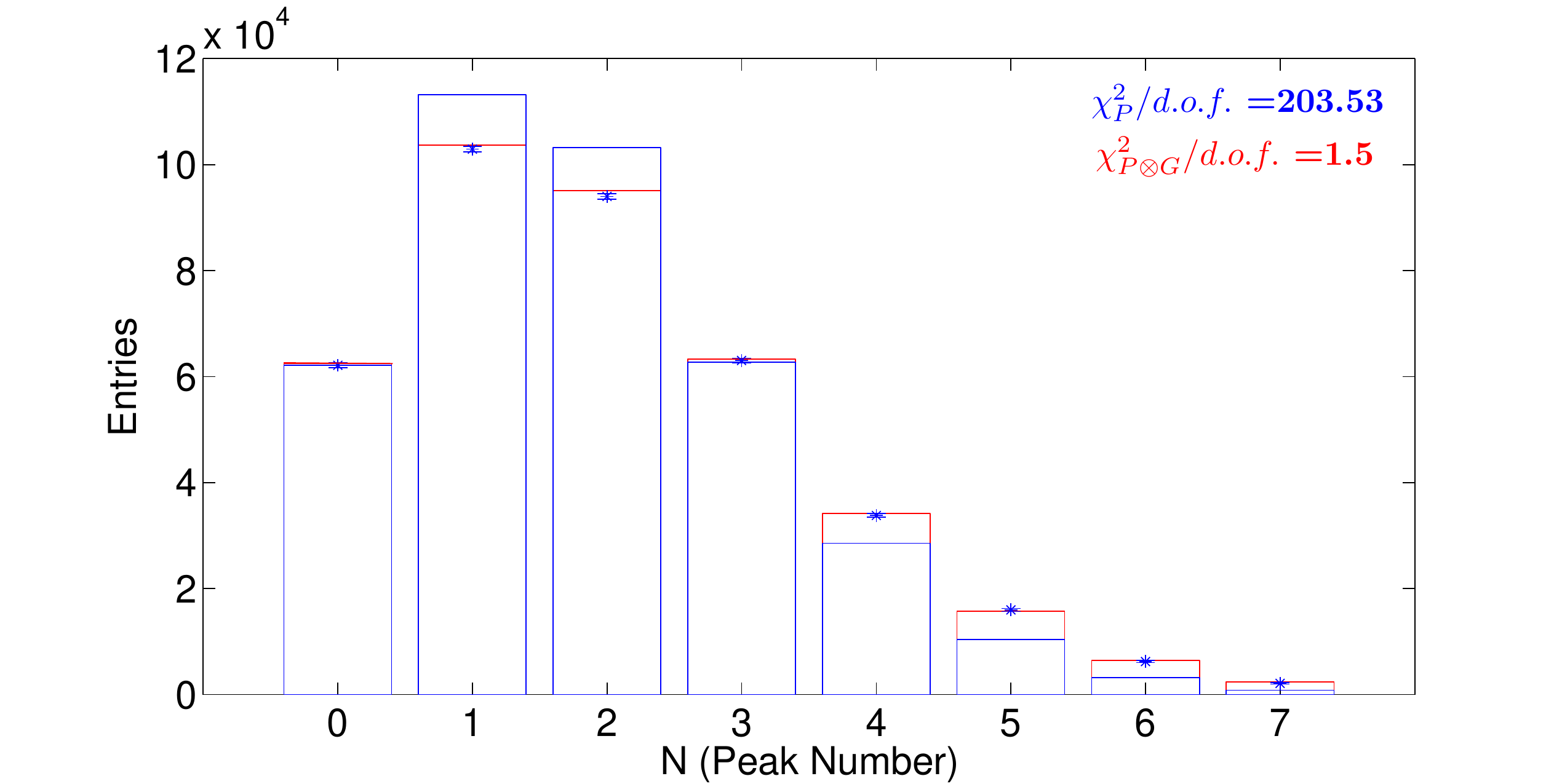}}
        \label{fig:AllRawData1}
        \end{subfigure}
        \vspace{-5pt}
        \caption{Results of the MGF procedure on the low and middle intensity, RUN1 and RUN2. The pure Poissonian model with mean value $\mu_{ZP}$ (blue) is compared to the $P\otimes G$ model (red).}
  \label{fig:FourPanels}
\vspace{-5pt}
\end{figure}

Results by the other recorded spectra are summarized in Table~\ref{tab:ConfrontingValues} and Fig.~\ref{fig:FourPanels} for the MGF procedure, confirming the validity of the compound Poissonian model and the need to account for detector effects to have a proper understanding of the phenomenon being investigated.

\begin{table}[h!]
\caption{Estimate of the mean number of fired cells with the $P\otimes G$ model using the RUN1 and RUN2 data-sets. Also in this case, the $P\otimes G$ model shows an agreement at the 99\% C.L.. The measured $\chi^2$ is 12.6 for the RUN1 and 12.0 for the RUN2 respectively.}
\label{tab:ConfrontingValues}
\centering
\scalebox{1.0}{
\begin{tabular}{p{2.6cm}ccc}
\hline\hline\\
                                          &   $\mu^0$	       &&	$\mu_{ZP}$ \\\\
                                          & $0.97 \pm 0.01$    &&	$0.985 \pm 0.002$ \\
Mean Value of the Poissonian distribution &&&\\
                                          & $1.82 \pm 0.01$    &&	$1.823 \pm 0.004$ \\ \\\hline \\
                                          &   $\mu_{MI}$	   &&	$\mu^0/(1-\epsilon_{XT}^0)$ \\\\
                                          & $1.080 \pm 0.002$      &&  $1.08 \pm 0.01$ \\
Mean Number of Fired Cells&&&\\
                                          & $1.994 \pm 0.003$      &&  $1.99 \pm 0.01$ \\
                  \\
 \hline\hline
    \end{tabular}    }
\end{table}

\section{Conclusion}

Instruments and methods for the investigation of the statistical properties of the light emitted by an incoherent source have been developed and validated. The experimental set-up is based on Silicon Photomultipliers, state-of-the art light detectors, embedded into a flexible, modular, easy-to-use kit. Methods fold the characteristics of the emitted light and the detector response, with an increasing level of refinement.  The model development allows to address advanced topics in statistics and data analysis techniques, targeted to master students in Physics or Engineering.\\

\section*{Acknowledgment}

The research reported here has been developed in the framework of a Joint Development Project between CAEN S.p.A. and Universita' degli Studi dell'Insubria. The analysis code developed by the authors in MATLAB\textsuperscript\textregistered  is made available to interested users.\\

\newpage

\ifCLASSOPTIONcaptionsoff
  \newpage
\fi

\newpage

\section*{Appendix A}\label{sec:A}

In this appendix it is demonstrated that the convolution of a Poissonian distribution of mean value $\lambda$ \eqref{eq:PoissonPDF} and a Binomial probability $\eta$ \eqref{eq:binomial} results in a Poissonian distribution of mean value $\lambda\eta$:\\  

\begin{itemize}
  \item{ Multiplying and dividing by $\eta^n$ each element in the series, Eq.~\ref{eq:phel} can be written as: 

$$
P_{d,el} = \sum_{n=d}^\infty B_{d,n}(\eta)P_{n,ph}(\lambda)
$$
$$
= \sum_{n=d}^\infty \frac{(\lambda\eta)^n \eta^{d-n}(1-\eta)^{n-d}e^{-\lambda}}{d!(n-d)!}.
$$}\\

\item{Hence, defining  $n-d=z$:

$$
P_{d,el}= \sum_{z=0}^\infty (\lambda\eta)^{d+z} \left( \frac{1-\eta}{\eta}\right)^z \frac{e^{-\lambda}}{d!z!} =
$$

$$
= \frac{(\lambda\eta)^d e^{-\lambda}}{d!}\times \sum_{z=0}^\infty \frac{(\lambda\eta)^z}{z!}\left(\frac{1-\eta}{\eta}\right)^z
$$
$$
= \frac{e^{-\lambda}(\lambda\eta)^d}{d!} \times \sum_{z=0}^\infty \frac{(\lambda - \lambda\eta)^z}{z!}.
$$}\\
\item{The series actually corresponds to the Taylor expansion of $e^{\lambda-\lambda\eta}$, so that:

$$
P_{d,el} = \sum_{n=d}^\infty B_{d,n}(\eta)P_{n,ph}(\lambda) =
$$
$$
\frac{e^{-\lambda\eta}(\lambda\eta)^d}{d!}
$$}\vspace{5pt}
\end{itemize}

\section*{Appendix B}\label{sec:B}

This appendix is dedicated to the demonstration of the relations for the probability density function \eqref{eq:pg}, the mean value and variance \eqref{eq:pgmv} of the total fired cell number $m$ assuming that every primary events can generate a unique infinite chain of secondary pulses.

This purpose is pursued by applying the probability generating function definition and properties.

For a discrete random variable $\phi$, the generating function is defined as:

\begin{equation*}
  \tilde{\Phi}(s)=\sum_{i=0}^{\infty}P(\phi=i)\times s^i .
\end{equation*}

The probability distribution function, the mean and the variance of the random variable $\phi$ can be calculated as:

\begin{equation}\label{eq:prob}
  \Phi(\phi=m)=\frac{1}{m!} \times \frac{\text{d}^m\Phi}{\text{d}s^m} \bigg|_{0}
  \end{equation}
\begin{equation}\label{eq:mean}
 \bar{m}_\Phi=\Phi(1)  
\end{equation}
\begin{equation}\label{eq:sigma}
    \sigma^2_\Phi=\Phi(1)^{\prime\prime}+\Phi(1)^{\prime}-{[\Phi(1)^{\prime}]}^2 .
\vspace{5pt}
\end{equation} 

The random variable considered here is the total number of detected pulses, $m$. Because it is defined by a sum of discrete random variables, its generating function is the composition of the pure Poisson distribution generating function:
\\
\begin{equation*}
  \tilde{P}(s)= e^{\mu(s-1)}
\end{equation*}

and of the geometric distribution generating function:
\\
\begin{equation*}
\begin{split}
\tilde{G}(s) & = \sum_{i=1}^{\infty}P(g=i-1)\times s^i\\
& = \sum_{i=1}^{\infty} {\epsilon_{XT}}^{i-1} \times (1-\epsilon_{XT}) \times s^i\\
& = \frac{(1-\epsilon_{XT})s}{1-\epsilon_{XT}s} .
\end{split}
 \end{equation*}
\\
Finally, the analytical expression of the generating function for the total number of fired cells result to be:
\\
\begin{equation*}
\begin{split}
  \tilde{P}\circ \tilde{G}& =\tilde{P}(\tilde{G}(s)) \\
  & = e^{\mu(\tilde{G}(s)-1)} \\
  & = e^{\mu\big(\frac{s-1}{1-\epsilon_{XT}s}\big)}.
  \end{split}
\end{equation*}
\\
Using the relation \eqref{eq:prob} it is possible to derive the probabilities to detect an arbitrary number of total pulses. For 0, 1 and 2 events the result is:
\\
\begin{equation*}
  P\otimes G(0)= e^{-\mu} ,
\end{equation*}
\begin{equation*}
  P\otimes G(1)= e^{-\mu}\mu (1-\epsilon_{XT}) ,
\end{equation*}
\begin{equation*}
  P\otimes G(2)= e^{-\mu} \left[\mu(1-\epsilon_{XT})\epsilon_{XT} + \frac{\mu^2(1-\epsilon_{XT})^2}{2}\right] .
\end{equation*}
\\
An analysis of these expressions lead to the compact and general formula reported in \eqref{eq:pg}, which refers to the compound Poisson distribution and is valid for $m$= 0, 1, 2,~...~. In addition, applying the properties \eqref{eq:mean} and \eqref{eq:sigma} at $\tilde{P}\circ \tilde{G}$, it is possible to obtain the mean value and the variance of the distribution of the total number of fired cells, as expressed by the relations in \eqref{eq:pgmv}.\\

\section*{Appendix C: the covariance ellipse}\label{sec:C}

In this appendix the confidence region of two variables is demonstrated to assume the shape of an ellipse. Moreover, the relation between the parameters describing the ellipse, the standard deviation of the variables and their correlation is established. 

The joint probability density of two variables $x^T$=$[ x_1, x_2]$ gaussian distributed may be written as:

  \begin{equation}\label{eq:joint}
    P(x)=k\cdot \text{exp}\Bigr\{-\frac{1}{2}(x-\mu)^T C^{-1}(x-\mu)\Bigl\} ,
      \end{equation}

where $k$ is a normalization constant, $\mu^T=[\mu_1 \mu_2]$ is the vector of the mean values of $x$ and $C$ is the covariance matrix:

\begin{equation*}
  C=E\{(x-\mu)(x-\mu)^T\}=\begin{bmatrix}
    \sigma_1^2 & \sigma_{12}\\
    \sigma_{21}&\sigma_2^2
  \end{bmatrix}.
\end{equation*} 

The diagonal elements of $C$ are the variances of the variables $x_i$ and the off-diagonal elements represent their covariance, which can be expressed as:

\begin{equation*}
  \sigma_{12}= \rho\sigma_1\sigma_2 ,
\end{equation*}     

where $\rho$ is the correlation coefficient. 

Curves of constant probability are determined by requiring the exponent of the equation \eqref{eq:joint} to be constant:

\begin{equation}\label{eq:ell}
  (x-\mu)^T C^{-1}(x-\mu)=c 
\end{equation}
\begin{equation*}  
\frac{(x_1-\mu_1)^2}{\sigma_1^2} -2\rho\frac{(x_1-\mu_1)}{\sigma_1}\frac{(x_2-\mu_2)}{\sigma_2} + \frac{(x_2-\mu_2)^2}{\sigma_2^2}=c' ,
\end{equation*}

where $c'=c(1-\rho^2)$. This equation represents an ellipse with the center located at ($\mu_1$, $\mu_2$) and the semi-axes placed at an angle $\theta$ with respect to the $x_1$, $x_2$ axes. 

As shown in the folllowing, the equation \eqref{eq:ell} can be re-written as a sum of squares of two stochastically independent variables, which results to be $\chi^2$ distributed with two degrees of freedom: 

\begin{equation}\label{eq:ellcan}
  \frac{\xi_1^2}{a^2}+ \frac{\xi_2^2}{b^2}= \chi^2 . 
\end{equation}

This relation describes an ellipse centered in the origin of the reference sistem and with the semi-axes of lenght $a$, $b$ parallel to the $\xi_1$,$\xi_2$ axes. 

As a first step, the origin of the reference system is translated in the center of the ellipse, resulting in equation:

 \begin{equation}\label{eq:ell2}
   \tilde{x}^T C^{-1}\tilde{x}=c ,
 \end{equation}

where $\tilde{x}= x-\mu$.

As a second step, axes are rotated in order to coincide with the ($\xi_1$, $\xi_2$) reference sistem by the transformation:

\begin{equation*}
 \tilde{x}=Q^T\xi , 
\end{equation*}
 
where  

\begin{equation*}
  Q = \begin{bmatrix}
    \cos \theta & \sin \theta \\
    -\sin \theta &\cos \theta
  \end{bmatrix} .
\end{equation*}

As a consequence, equation \eqref{eq:ell2} is turned to the form

 \begin{equation*}
   \xi^T Q C^{-1} Q^T \xi=c ,
 \end{equation*}

corresponding to the equation \eqref{eq:ellcan} as long as

\begin{equation*}
  Q C^{-1} Q^T =\begin{bmatrix}
    \frac{1}{a^2} & 0\\
    0&\frac{1}{b^2}
  \end{bmatrix} ,
\end{equation*}

or, equivalently,

\begin{equation*}
  Q C Q^T =\begin{bmatrix}
    a^2 & 0\\
    0&b^2
  \end{bmatrix} .
\end{equation*}

The vector of the mean values and the cvariance matrix of $\xi$ results to be:

\begin{equation}\label{eq:cov}
\begin{split}
 \mu_\xi&=E\{\xi\}=Q E\{x\}=Q\mu\\
  C_\xi&=E\{(\xi-\mu_\xi)(\xi-\mu_\xi)^T\} \\
  &=Q E\{(x-\mu)(x-\mu)^T\}Q^T \\
  &=Q C Q^T
\end{split}
\end{equation}

So it can be noticed that the eigenvalues of the covariance matrix $C_\xi$ correspond to the squared semi-axes of the canonical ellipse \eqref{eq:ellcan}. 

Because of the symmetry of the covariance matrix, $C$ can be diagonalized exploiting its decomposition in eigenvalues and eigenvectors:

\begin{equation*}
  C=U\Lambda U^T ,
\end{equation*}  
 
where $\Lambda$ is the diagonal matrix of eigenvalues and $U$ is the rotation matrix constitued by eigenvectors. Comparing this formula with the expression \eqref{eq:cov} and using the properties of the rotation matrix ($QQ^T=Q^TQ=I$, det$Q=1$) it can be inferred that:

\begin{equation*}
  U=Q^T \qquad \qquad \Lambda=C_\xi .
\end{equation*} 

As a consequence, the eigenvalues of $C$ can be obtained through the quadratic equation:

\begin{equation*}
  \text{det}(C-\lambda I)=0 ,
\end{equation*}

whose solutions are:

\begin{equation*}
  \lambda_{1,2}= \frac{1}{2}\Bigl[ (\sigma_1^2+\sigma_2^2) \pm \sqrt{(\sigma_1^2+\sigma_2^2)^2-4\sigma_1^2\sigma_2^2(1-\rho)} \Bigr] .
\end{equation*}
\\
The lenghts of the ellipse semi-axes result to be the square root of the eigenvalues multiplied by the two degrees of freedom $\chi^2$ value:

\begin{equation}\label{eq:semiaxes}
  a=\sqrt{\chi^2\lambda_1} \qquad \qquad b=\sqrt{\chi^2\lambda_2} .
\end{equation}

The eigenvectors of $C$ can be found with the following equation:

\begin{equation*}
  (C-\lambda_iI)u_i=0 , \quad  \text{with} \quad i=1,2. 
\end{equation*}

For $i=1$:

\begin{equation*}
  \begin{bmatrix}
    \sigma_1^2-\lambda_1 & \rho\sigma_1\sigma_2 \\
    \rho\sigma_1\sigma_2 & \sigma_2^2-\lambda_1
  \end{bmatrix}
  \begin{bmatrix}
    u_{1,1} \\
    u_{1,2}
  \end{bmatrix}
  =0 ,
\end{equation*} 

and the solution is:

\begin{equation*}
  u_1=\alpha_1 \begin{bmatrix}
    -\rho\sigma_1\sigma_2\\
    \sigma_1^2-\lambda_1
  \end{bmatrix} ,
\end{equation*}

where $\alpha_1$ is a normalization constant. In the case of $i=2$:

\begin{equation*}
  \begin{bmatrix}
    \sigma_1^2-\lambda_2 & \rho\sigma_1\sigma_2 \\
    \rho\sigma_1\sigma_2 & \sigma_2^2-\lambda_2
  \end{bmatrix}
  \begin{bmatrix}
    u_{2,1} \\
    u_{2,2}
  \end{bmatrix}
  =0 ,
\end{equation*} 

and the solution is:

\begin{equation*}
  u_2=\alpha_2 \begin{bmatrix}
     \sigma_2^2-\lambda_2\\
     -\rho\sigma_1\sigma_2
  \end{bmatrix} ,
\end{equation*}

where $\alpha_2$ is the normalization constant. Using the eigenvalues definition, it can be proved that $\sigma_2^2 - \lambda_2 = -(\sigma_1^2 - \lambda_1)$. As a result, the $U$ matrix turns out to be equal to $Q^T$, with $\cos\theta=-\rho\sigma_1\sigma_2$ and $\sin\theta=\sigma_1^2-\lambda_1$. From these identities it is possibile to calculate the angle $\theta$ between the ellipse axis, which lies on $\xi_i$, and the $x_i$ axis:

\begin{equation*}
  \tan\theta= - \frac{\sigma_1^2-\lambda_1}{\rho\sigma_1\sigma_2} .
\end{equation*}

As $\theta$ belongs to the range [-$\pi$/2, $\pi$/2] and the above expression is quite complex, it is more convenient to estimate the $\tan2\theta$:

\begin{equation}\label{eq:tan2}
  \tan2\theta= \frac{2\tan\theta_1}{1-\tan^2\theta_1} = \frac{2 \rho\sigma_1\sigma_2}{\sigma_1^2-\sigma_2^2} .
\end{equation} 

The angle $\theta$ measures the rotation which brings the ($x_1$, $x_2$) coordinate system in the ($\xi_1$, $\xi_2$) reference system, which represent the rotation undergone by the ellipse. The rotation matrix $Q$ has been completely determined and the ellipse has been entirely defined. 

The covariance ellipse of the bivariate normal distribution assumes a particular importance when $\chi^2=1$ and its features can be analyzed in two extreme cases: \\

\begin{itemize}
\item{if the variables are not correlated ($\rho~=~0$), then $\theta~=~0$, $a=\sigma_1$ and $b=\sigma_2$, which means that the ellipse axes are parallel to $x_i$ and equal to the variable standard deviations,} 
\item{if the correlation is maximum ($\rho=\pm1$), then the ellipse degenerates into a straight line of lenght $a=\sqrt{\sigma_1^2+\sigma_2^2}$ (in fact $b=0$).}\\ 
\end{itemize}

In all the intermediate cases the ellipse is inscribed in a rectangle of center ($\mu_1,\mu_2$) and sides $2\sigma_1$ and $2\sigma_2$. The projections on the $x_i$ axes of the four intersection points between the ellipse and the rectangle represent the 68\% confidence interval for the parameter centered in the mean value $\mu_i$. \\

All these characteristics of the covariance ellipse can be demonstrated exploiting the conic equations. 
The general quadratic equation:

\begin{equation}\label{eq:second}
  Ax_1^2 + Bx_1x_2 + Cx_2^2 + Dx_1 +Ex_2 + F = 0
\end{equation}

represents the canonical ellipse if $B=0$ and $AC>0$. It is always possible to find a new coordinate system, rotated by an angle $\theta$ with respect to the $x_i$ axes, in which the equation does not involve the mixed variable product. Calling $\xi_i$ the new set of axis, the $x_i$ variables can be expressed as:

\begin{equation*}
  x_1=\xi_1\cos\theta -\xi_2\sin\theta \quad x_2=\xi_1\sin\theta +\xi_2\cos\theta .
\end{equation*}

Substituing these relations in \eqref{eq:second} and collecting the similar terms a new equation in $\xi_i$ can be obtained:

\begin{equation}\label{eq:second2}
\begin{split}
  &\xi_1^2(A\cos^2\theta+B\cos\theta\sin\theta+C\sin^2\theta)+\\
  &\xi_1\xi_2(-2A\cos\theta\sin\theta +B(\cos^2\theta-\sin^2\theta)+2C\sin\theta\cos\theta)+\\
  &\xi_2^2(A\sin^2\theta-B\cos\theta\sin\theta+C\cos^2\theta)+\\
  &\xi_1(D\cos\theta+E\sin\theta) + \xi_2(-D\sin\theta+E\cos\theta)+F=0 .
  \end{split}
\end{equation} 
 
In order to eliminate the $\xi_1\xi_2$ term, the angle $\theta$ has to satisfy:

\begin{equation*}
  -2A\cos\theta\sin\theta +B(\cos^2\theta-\sin^2\theta)+2C\sin\theta\cos\theta =0.
\end{equation*} 

Simplifying the equation:

\begin{equation}\label{eq:abc}
\begin{split}
  2(A-C)\cos\theta\sin\theta&=B(\cos^2\theta-\sin^2\theta)\\
  \frac{2\sin\theta\cos\theta}{\cos^2\theta-\sin^2\theta}&=\frac{B}{A-C}\\
  \tan2\theta&=\frac{B}{A-C}
  \end{split}
\end{equation}

In the specific case corresponding to equation \eqref{eq:ell},
 
\begin{equation*}
 A=\frac{1}{\sigma_1^2} \quad B=-\frac{2\rho}{\sigma_1\sigma_2} \quad C=\frac{1}{\sigma_2^2} .
\end{equation*}

As a consequence the expression \eqref{eq:abc} assume the form of the relation \eqref{eq:tan2}. Finally, the coefficients of the second order variables in equation \eqref{eq:second2} have to be interpreted as the inverse square of the semi-axes lenghts. Replacing the definition of $A$, $B$ and $C$ and solving for $a$ and $b$ gives:

\begin{equation*}
  a= \sqrt{\frac{\sigma_1^2\sigma_2^2(1-\rho^2)}{\sigma_2^2\cos^2\theta - 2\rho\sigma_1\sigma_2\cos\theta\sin\theta + \sigma_1^2\sin^2\theta^2}}
\end{equation*}
\begin{equation*}
  b= \sqrt{\frac{\sigma_1^2\sigma_2^2(1-\rho^2)}{\sigma_2^2\sin^2\theta - 2\rho\sigma_1\sigma_2\cos\theta\sin\theta + \sigma_1^2\cos^2\theta^2}} .
\end{equation*}
\\
Expressing $\theta$ as a function of $\rho$ and $\sigma_i$ it is possible to obtain for the semi-axes the same definition as found previously in equation \eqref{eq:semiaxes}.
\end{document}